\documentclass[12pt]{article}
\usepackage{amsbsy,amsmath}
\usepackage{amsfonts}
\usepackage{amssymb}
\usepackage{amscd}
\usepackage{bbm}
\usepackage{fancybox}
\usepackage{cite}
\usepackage{amsmath,amsfonts,amsbsy}
\usepackage{pstricks,pst-node}
\usepackage[small,bf,hang]{caption2}
\usepackage{graphicx}
\usepackage{epsfig}
\usepackage{psfrag}
\usepackage{comment}

\usepackage[mathscr]{euscript}

\usepackage{float}

\psset{unit=1.3cm,linewidth=.5pt,radius=.2}  

\usepackage{multirow}                     
\usepackage{float}                          
\usepackage{lscape}                         
\usepackage{bm}

\newcommand{\be}{\begin{equation}}
\newcommand{\ee}{\end{equation}}
\newcommand{\bea}{\begin{eqnarray}}
\newcommand{\eea}{\end{eqnarray}}
\newcommand{\ba}{\begin{array}}
\newcommand{\ea}{\end{array}}

\def\HH{\mathcal{H}}
\def\LL{\mathcal{L}}
\def\hh{\mathfrak{h}}
\def\TT{\mathcal{T}}

\font\mybb=msbm10 at 12pt
\def\bb#1{\hbox{\mybb#1}}

\makeatletter \@addtoreset{equation}{section} \makeatother

\def\slashchar#1{\setbox0=\hbox{$#1$}           
   \dimen0=\wd0                                 
   \setbox1=\hbox{/} \dimen1=\wd1               
   \ifdim\dimen0>\dimen1                        
      \rlap{\hbox to \dimen0{\hfil/\hfil}}      
      #1                                        
   \else                                        
      \rlap{\hbox to \dimen1{\hfil$#1$\hfil}}   
      /                                         
   \fi}

\begin{document}

\begin{titlepage}

\begin{center}

\vskip 1.5cm

{\Large \bf Dualities of Self-Dual Nonlinear Electrodynamics}

\vskip 1cm

{\bf Jorge G.~Russo\,${}^{a,b}$ and  Paul K.~Townsend\,${}^c$} \\

\vskip 25pt

{\em $^a$  \hskip -.1truecm
\em Instituci\'o Catalana de Recerca i Estudis Avan\c{c}ats (ICREA),\\
Pg. Lluis Companys, 23, 08010 Barcelona, Spain.
 \vskip 5pt }

\vskip .4truecm

{\em $^b$  \hskip -.1truecm
\em Departament de F\' \i sica Cu\' antica i Astrof\'\i sica and Institut de Ci\`encies del Cosmos,\\ 
Universitat de Barcelona, Mart\'i Franqu\`es, 1, 08028
Barcelona, Spain.
 \vskip 5pt }
 
 \vskip .4truecm

{\em $^c$ \hskip -.1truecm
\em  Department of Applied Mathematics and Theoretical Physics,\\ Centre for Mathematical Sciences, University of Cambridge,\\
Wilberforce Road, Cambridge, CB3 0WA, U.K.\vskip 5pt }

\hskip 1cm

\noindent {\it e-mail:}  {\texttt jorge.russo@icrea.cat, pkt10@cam.ac.uk}

\end{center}

\vskip 0.5cm
\begin{center} {\bf ABSTRACT}\\[3ex]
\end{center}

\noindent

For any causal nonlinear electrodynamics theory that  is ``self-dual'' (electromagnetic $U(1)$-duality invariant), 
the Legendre-dual pair of Lagrangian and Hamiltonian densities $\{\LL,\HH\}$ are constructed from functions  
$\{\ell,\hh\}$ on $\bb{R}^+$ related to a particle-mechanics Lagrangian and Hamiltonian.
We show how a `duality' relating $\ell$ to $\hh$ implies that $\LL$ and $\HH$ are related by a simple map 
between appropriate pairs of variables.  We also discuss Born's  ``Legendre self-duality'' and  implications of a new 
``$\Phi$-parity'' duality. Our results are illustrated with many examples. 

\vfill

\end{titlepage}
\tableofcontents
\section{Introduction}

Nonlinear theories of electrodynamics (NLED) are generally defined \cite{Born:1933lls,Born:1934gh,Plebanski:1970zz,Boillat:1970gw} 
by means of a Lagrangian density function $\mathcal{L}(S,P)$ of the two Lorentz (pseudo)scalars 
\be\label{SPdefs}
S= \frac12\left(E^2 - B^2\right)  \, , \qquad P= {\bf E}\cdot {\bf B}\, , 
\ee
where $({\bf E},{\bf B})$ are the (electric, magnetic) 3-vector-field components of the abelian 2-form field strength $F=dA$ on Minkowski spacetime, and $(E,B)$ are their respective magnitudes. A feature of all NLED theories in this ``Plebanski'' class is that the degrees of freedom remain those of the free-field Maxwell theory, although superluminal signal propagation (and hence causality violation) is potentially possible, and the physical theories are those for which it is not possible. These features are shared by some NLED theories outside the Plebanski class, some of which are physical \cite{Bialynicki-Birula:1984daz,Bialynicki-Birula:1992rcm,Russo:2022qvz,Mezincescu:2023zny}, but they have no weak-field limit. In this paper we assume the existence of a (conformal) weak-field limit. 

We also focus on a class of NLED theories that share with the free-field Maxwell electrodynamics the property of electromagnetic duality invariance. In the Maxwell case this can be viewed as an invariance of the (source-free) Maxwell equations under any constant shift of the phase of the complex 3-vector field ${\bf E}+ i{\bf B}$. However, this definition applies only in Cartesian coordinate systems since ${\bf E}$ and ${\bf B}$ are 3-vectors in dual vector spaces. A better definition, which not only applies for any coordinate system but also generalises to nonlinear electrodynamics, is as an invariance of the Hamiltonian density under any constant phase shift of the complex 3-vector field \cite{Schrodinger:1935oqa,Gaillard:1981rj}
\be
{\bf D} + i{\bf B} \, , \qquad {\bf D} := \frac{\partial \mathcal{L}}{\partial {\bf E}}\, . 
\ee
Invariance of the field equations is then a consequence of the invariance of the 
Hamiltonian. For the Maxwell case, ${\bf D}={\bf E}$ in Cartesian coordinates, and 
we therefore recover the earlier definition. Following what has become standard terminology, we shall say that any NLED theory with this $U(1)$ symmetry is ``self-dual''.  

Within the Lagrangian formulation, the restriction to a self-dual theory is achieved by requiring the Lagrangian density to satisfy the following partial differential equation (PDE) \cite{Bialynicki-Birula:1984daz,Gibbons:1995cv,Gaillard:1997rt}:
\be\label{BB-SD}
P\left(\LL_S^2-\LL_P^2 -1\right) = 2S \LL_S\LL_P\, . 
\ee
The first example, excepting the free-field Maxwell case, was the Born-Infeld 
theory \cite{Born:1934gh}, although its self-duality was noticed by Schr\"odinger
a few years later \cite{Schrodinger:1935oqa}.  Since the re-appearance of Born-Infeld electrodynamics in the 1990s as (or as part of) an effective theory for open strings of string theories  \cite{Fradkin:1985qd,Bergshoeff:1987at,Leigh:1989jq,Tseytlin:1999dj} 
there has been a resurgence of interest in nonlinear electrodynamics. In particular, the possibility of a Born scale in electrodynamics is now taken seriously for its potential relevance to the physics of magnetars, e.g.  \cite{Pereira:2018mnn,Denisov:2003ba}, and to particle physics experiments at future colliders \cite{Ellis:2022uxv}. 

One motivation for our focus on self-dual NLED theories is that many of the special properties of Born-Infeld are consequences of its self-duality. Another motivation is the recent result that strong-field causality is implied by weak-field causality for all self-dual NLED with a weak-field limit \cite{Russo:2024llm}. We elaborate below on the significance of this fact, and  one purpose of this paper is to provide more details of the results of \cite{Russo:2024llm}. 

Another purpose is to expand the results of \cite{Russo:2024llm} to include the Hamiltonian formulation. As this is equivalent to the Lagrangian formulation for all causal NLED theories, we did not expect surprises.  However, various additional remarkable properties of self-dual theories emerge from the conjunction of the Lagrangian and Hamiltonian formulations. For the remainder of this Introduction we provide the necessary background and a sketch of our main new results. 

The self-duality PDE \eqref{BB-SD} can be simplified by expressing $\LL$ as a function of 
$S$ and 
\be\label{defPhi}
\Phi := \sqrt{S^2+P^2}\, . 
\ee
This is possible only if $\LL$ preserves parity since both $S$ and $\Phi$ are parity even whereas $P$ is parity odd, but this restriction is not a limitation for self-dual NLED because self-duality implies parity, for a reason to be explained below. The self-duality
PDE for $\LL(S,\Phi)$ is \cite{Gaillard:1997rt}
\be\label{PDEL}
\LL_S^2 - \LL_\Phi^2 =1\, . 
\ee
For some purposes it is convenient to use the alternative independent variables
\be\label{UVeq}
U=\frac12(\Phi-S)\ ,\qquad V=\frac12(\Phi+S)\, . 
\ee
Notice that $(U,V)$ as defined are both non-negative because $\Phi\ge |S|$. This implies that the `physical' values of $(U,V)$ are restricted to the positive quadrant in the $(U,V)$-plane. The self-duality PDE for $\LL(U,V)$ is \cite{Gibbons:1995cv}
\be\label{GR}
\mathcal{L}_U \mathcal{L}_V = -1\, .     
\ee
The general solution to this equation in the positive $(U,V)$-quadrant, in terms of the boundary function $\ell(V) :=\LL(0,V)$, is \cite{C&H}
\be\label{gensol}
\mathcal{L}=\ell (\tau)-\frac{2U}{\dot\ell(\tau)}\ , 
\qquad  \tau=V+\frac{U}{\dot\ell^2(\tau)}\, , 
\ee
where
\be
\dot\ell(\tau) = \frac{d\ell(\tau)}{d\tau} >0\, . 
\ee
We shall call this the Courant-Hilbert (CH) solution, and $\ell(\tau)$ a ``CH-function''. Notice that $\tau\ge0$ by definition, with equality only for $U=V=0$. The choice $\ell(\tau)=\tau$ yields the free-field Maxwell case.

To verify that \eqref{gensol} solves \eqref{GR}, we may take the differential of both sides of both equations of \eqref{gensol}. The resulting pair of equations for the differentials may then be solved for $d\LL$ and $d\tau$ in terms of $dU$ and $dV$. The result for $d\LL$ implies 
\be\label{partials1}
\LL_V  = \dot\ell \, , \qquad \LL_U = -{\dot\ell}^{-1} \, , 
\ee
which confirms that $\LL_U \LL_V=-1$. The result for $d\tau$ is 
\be\label{intersect}
Gd\tau = \dot\ell(dU + \dot\ell^2 dV) \,, 
\ee
where
\be\label{defG}
G:= \dot\ell^3 + 2 \ddot\ell U\, .
\ee
The main implications of \eqref{intersect} were briefly discussed in \cite{Russo:2024llm} and we review this, with more detail, in the following section. 

Within the Plebanski NLED class, the necessary and sufficient conditions for causality were found in \cite{Schellstede:2016zue}, subject to an assumption about the domain of the function $\mathcal{L}(S,P)$ that can be interpreted physically as the existence of a weak-field limit.  These conditions can be separated into two sets according to whether a violation is possible (generically) for weak fields, or only for strong fields. The former set are 
equivalent to convexity of the function $\mathcal{L}(S,P)$, which are also the conditions for convexity of $\mathcal{L}$ as a function of ${\bf E}$\cite{Bandos:2021rqy}. The remaining (strong-field) causality condition was provided with some intuition and an alternative derivation in \cite{Russo:2024kto}.   For {\sl self-dual} NLED theories with a weak-field limit, we showed in \cite{Russo:2024llm}  that all these causality conditions reduce  to the following simple inequalities to be satisfied by derivatives of the CH-function $\ell(\tau)$:
\be\label{causal-L}
\dot\ell \ge 1 \, , \qquad \ddot \ell\ge0 \, . 
\ee
Apart from its simplicity, this result is remarkable because there was no {\it a priori} reason to suppose that the 
causality conditions on $\ell$  would be independent of $(U,V)$. Notice that the condition $\ddot\ell \ge0$ tells us that $\ell(\tau)$ is a convex function. 

The assumption of a weak-field limit can also be expressed in terms of the function $\ell(\tau)$;  it is the statement that $\ell(\tau)$ should have a Taylor-series expansion in $\tau$.  Omitting the constant term in this expansion on the grounds that it is irrelevant to the NLED dynamics; we have
\be\label{series}
\ell(\tau) = e^\gamma \tau + \mathcal{O}(\tau^2)\, , 
\ee
for some dimensionless constant $\gamma$, which must be non-negative in order to satisfy the causality condition $\dot\ell\ge1$ in the weak-field limit. In this limit $\ell(\tau) = e^\gamma\tau$, which yields the free-field Maxwell theory for $\gamma=0$ and ModMax (the ``modified Maxwell'' theory \cite{Bandos:2020jsw}) for $\gamma>0$. Both are conformal because the conformality condition for self-dual NLED is equivalent to degree-1 homogeneity of  $\ell(\tau)$, as we show in section \ref{sec:SDH}. Another feature of the existence of a weak-field expansion is that $\ell(\tau)$ defines a function not only for $\tau\ge0$ (which is all that is relevant to the CH solution of the self-duality PDE) but also for $\tau<0$, at least in some neighbourhood of $\tau=0$. We shall see later the significance of this fact. 

A feature of the CH equations \eqref{gensol} is that many simple functions $\ell(\tau)$ satisfying \eqref{causal-L} allow $\LL(U,V)$ to be found analytically, leading to explicit Lagrangian densities for a variety of causal self-dual NLED theories. These include Born-Infeld \cite{Born:1934gh} and its Mod-Max-type generalisation \cite{Bandos:2020jsw,Bandos:2020hgy} that we call, for brevity, ``ModMaxBorn''. Other examples were given in \cite{Russo:2024llm} and more will be given here.  

We shall expand on the results of \cite{Russo:2024llm} in the following section but, as stated above, our main purpose is to explore the Hamiltonian formulation for self-dual NLED. An advantage of this formulation is that self-duality can be implemented simply by restricting the Hamiltonian density $\mathcal{H}({\bf D},{\bf B})$ to be a function of the 
two duality-invariant rotation scalars 
\be\label{sandp}
s= \frac12 \left(D^2 + B^2\right) \, , \qquad  p = |{\bf D}\times {\bf B}|\,  \qquad (D= |{\bf D}|).  
\ee 
As both $s$ and $p$ are parity-even (parity flips the sign of ${\bf D}$) it follows that any function $\HH(s,p)$ is both duality {\sl and} parity invariant, and hence that {\sl all self-dual NLED theories preserve parity}, as claimed above. This result was proved (although not stated) in \cite{Kuzenko:2000uh}, is implicit in \cite{Ivanov:2003uj}, and may be known to others, but it is possibly not appreciated how obvious it becomes in the Hamiltonian formulation. 

A disadvantage of the Hamiltonian formulation is that Lorentz invariance is not manifest.
The condition for a generic, and not necessarily duality-invariant, Hamiltonian density to define a Lorentz invariant NLED was found in \cite{Bialynicki-Birula:1984daz}. Here we need this condition for functions of $(s,p)$ only, and if we trade these variables for $s$ and 
\be\label{defvp}
\varphi := \sqrt{s^2-p^2}\, ,   
\ee
then the Lorentz invariance condition for $\HH(s,\varphi)$ is the PDE
\be\label{PDEH}
\HH_s^2 -\HH_\varphi^2 =1\, ,  
\ee
which is formally identical to the Lagrangian self-duality PDE of \eqref{PDEL}. 
 
For some purposes it is convenient to use the new independent variables\footnote{These differ from the definitions of $(u,v)$ in \cite{Bandos:2020jsw} by the exchange $u\leftrightarrow v$, which facilitates comparison between the Lagrangian and Hamiltonian formulations of self-dual NLED.} 
\be\label{uandv}
u=\frac12 \left(s-\varphi\right)\ ,\qquad 
v=\frac12 \left(s+\varphi\right)\, , 
\ee
Notice that $(u,v)$ are both non-negative, and that $v\ge u$, so the `physical' region in the 
$(u,v)$-plane is the region of the positive quadrant bounded by $u=0$ and $v=u$. 
The condition for $\HH(u,v)$ to define a Lorentz invariant NLED is \cite{Bandos:2020hgy}
\be\label{GR-H}
\mathcal{H}_u \mathcal{H}_v =1\, .  
\ee
This is mathematically equivalent to \eqref{GR} since the sign on the right-hand side can be changed  by using $(-u,v)$ instead of $(u,v)$ as the independent variables, and the general solution for $\mathcal{H}$ is then formally the same as the solution of \eqref{gensol} for $\mathcal{L}$. However, we shall use the variables $(u,v)$ as defined above because they are both non-negative. Notice that \eqref{GR-H} has the solution 
\be\label{BB}
\HH(u,v) = \sqrt{4uv} =p\, , 
\ee 
which defines the conformal Bilaynicki-Birula electrodynamics \cite{Bialynicki-Birula:1984daz}. There is no analogous solution of \eqref{GR} because of the different sign on the right-hand side. 

All other solutions of \eqref{GR-H}, expressed in terms of the boundary function $\HH(0,v)=\frak{h}(v)$, are given by 
\be
\label{generalh}
\HH = \hh(\sigma)+\frac{2u}{\hh'(\sigma)}\ ,\qquad \sigma=v-\frac{u}{\left[\hh'(\sigma)\right]^2} \qquad (\hh^\prime>0), 
\ee
where $\hh(\sigma)$ is a new CH-function analogous to $\ell(\tau)$. 

Corresponding to every causal NLED defined by a function $\mathcal{L}(U,V)$ there is a Hamiltonian density function $\mathcal{H}(u,v)$ and the two are related by a Legendre transform. This is because convexity of $\mathcal{L}$ (as a function of ${\bf E}$) implies convexity of $\mathcal{H}$ (as a function of ${\bf D}$) and this implies that the Legendre transform is an 
involution, although ``strict'' convexity (non-zero Hessian determinant) is needed to apply this theorem to the Plebanski class of NLED theories.  For self-dual theories a corollary of this correspondence is that the functions $\ell(\tau)$ and $\hh(\sigma)$ must be related in some way that allows one to be found from the other. What we find is that the following functions are Legendre transforms of each other:
\be\label{L&H}
L(\sqrt{2\tau}) = \ell(\tau)\, , \qquad H(\sqrt{2\sigma}) = \hh(\sigma)\, . 
\ee
In other words, the functions $\ell$ and $\hh $ are related by a Legendre transform\footnote{As we explain later, this requires $\sigma\geq 0$.}
but in terms of the new variables $\sqrt{2\tau}$ and $\sqrt{2\sigma}$.
Our choice of notation is motivated by the fact that the functions $L$ and $H$ can be interpreted as the Lagrangian and Hamiltonian of a particle mechanics model associated to the NLED defined by the Lagrangian and Hamiltonian densities $\LL$ and $\HH$. This was a motivating analogy for Born's original NLED theory, and a correspondence between Born-Infeld and the massive relativistic particle is a consequence of T-duality for the effective worldvolume field theories of D-branes (see e.g. \cite{Bergshoeff:1996cy,Green:1996bh}). However, the
correspondence applies more generally, as we discuss in section \ref{sec:SPD}.

Our Hamiltonian results for self-dual NLED theories allow us to `translate' the causality conditions on $\ell(\tau)$ to corresponding causality conditions on $\hh(\sigma)$. As we shall see, these are 
\be\label{hprime}
0< \hh^\prime(\sigma) \le1 \, , \qquad \hh^\prime{}^\prime(\sigma)  \le 0\, , 
\ee
{\sl and} 
\be\label{hprime'}
\hh^\prime(\sigma) + 2\sigma \hh^\prime{}^\prime(\sigma) >0\, . 
\ee
This last condition  is equivalent to strict convexity of the function $H(\sqrt{2\sigma})$, which is required for its interpretation as the Legendre dual of  $L(\sqrt{2\tau})$, which is in turn required for the interpretation of $\mathcal{H}$ as the Legendre dual of $\mathcal{L}$. Notice that $\hh(\sigma)$ is required to be a concave function ($\hh^\prime{}^\prime\le 0$) in contrast to the convexity condition ($\ddot\ell\ge0$) on $\ell(\tau)$. 

Surprisingly, the function $\hh$ can be used to directly construct
not only the Hamiltonian density but also the Lagrangian density, again via the 
Courant-Hilbert solution but now for boundary conditions at $V=0$ rather than $U=0$. 
Since $U$ and $V$ are exchanged by an exchange of ${\bf E}$ and ${\bf B}$, this is a 
type of electromagnetic duality, which is indirectly equivalent to a Legendre duality. 
As we shall see, this fact implies a remarkably simple relation between the Lagrangian and 
Hamiltonian densities of any self-dual NLED. For example, given the Lagrangian density 
in the form $\LL(S,\Phi)$ the Hamiltonian density in the form $\HH(s,\varphi)$ can be found from the following procedure:
\be\label{magic}
\boxed{\LL(S,\Phi) \ \longrightarrow \  -\LL(-s,\varphi)= \HH(s,\varphi)}\ . 
\ee
This allows us to find $\HH$ from $\LL$ {\sl without the need for a Legendre transform}! 
For the free-field Maxwell case, we have
\be
\LL=S \ \longrightarrow \ -(-s) = \HH \quad \Rightarrow \ \HH =s\, . 
\ee

This result suggests that $\ell$ and $\hh$ must be similarly related since they determine $\LL$ and $\HH$. We find, in some cases, that there is indeed a very simple relation between these two CH-functions, but the 
general case requires consideration of what we call the ``$\Phi$-parity'' (equivalently, ``$\varphi$-parity'') dual NLED defined by 
\be
\hat\LL(S, \Phi) := \LL(S,-\Phi)\, , \qquad 
\hat\HH(s,\varphi) := \HH(s,-\varphi) \, . 
\ee
In some cases, such as Born-Infeld, $\hat\LL=\LL$. For this $\Phi$-parity invariant subset of self-dual NLED we find that 
\be\label{lhx}
\ell(x) + \hh(-x) =0 \, , \qquad x\in \bb{R}\, , \qquad \ \ (\hat\LL = \LL). 
\ee
As mentioned above, only the values of $\ell(x)$ for $x\ge0$ are relevant to the CH solution for $\LL(U,V)$, but $\ell(x)$ is defined for $x<0$ if there is a weak field limit (which we are assuming here). We now see that for any $\Phi$-parity invariant NLED (with a weak-field limit) the CH function $\ell(x)$ for $x\le0$ determines the other CH function $\hh$. It remains true, of course, that $\ell$ and $\hh$ are related by their relation to the Legendre dual pair of functions $\{L,H\}$, but no Legendre transform is needed to find one from the other! 

More generally, $\hat\LL\ne\LL$ and $\hat\LL$ is associated with a pair of CH-functions $\{\hat\ell,\hat\hh\}$ that differ from $\{\ell,\hh\}$. For these cases we find that  $\{\ell,\hat\ell\}$ are related to $\{\hh,\hat\hh\}$ by a pair of relations similar to \eqref{lhx} but intertwined by $\Phi$-parity. 

An obvious question is whether there is a simple characterisation, in terms of restrictions on the CH functions $\{\ell,\hh\}$, of the subclass of self-dual NLED theories that are also $\Phi$-parity invariant. There is, but it involves an alternative solution of the self-duality PDE, also given by Courant and Hilbert \cite{C&H}, in terms of a function $\omega(x)$ of a positive variable $x$. The $\Phi$-parity invariant self-dual theories are those for which $\omega(x)$ is invariant under $x\to 1/x$, and for these theories
we show that $\omega(x)$ is the Legendre-dual of $\ell(\tau)$ {\sl with respect to $\tau$} (rather than $\sqrt{2\tau})$.  Born-Infeld corresponds to the choice of a linear function of  $(x+1/x)$. 

Another topic that we discuss is ``Legendre self-duality'', which has no direct 
connection to the topics described above, but could potentially be confused 
with them.  The Hamiltonian density $\HH$ is the Legendre transform of the Lagrangian density $\LL$ with respect to ${\bf E}$. If we now take the Legendre transform of $\HH$ with respect to ${\bf B}$ we arrive at a `dual' Lagrangian density $\tilde\LL$, which is a function of Lorentz scalars constructed from the 
Legendre-duals of $({\bf E}, {\bf B})$. It was noticed by Born, for Born-Infeld, that $\LL$ and $\tilde\LL$ are the same function\footnote{Born refers, confusingly,  to the dual Lagrangian as the ``Hamiltonian'', with the recognition that this is an abuse of terminology.} of their respective Lorentz scalars \cite{Born:1937drv}. Much later, it was shown by Gaillard and Zumino \cite{Gaillard:1997rt} that any (electromagnetically) self-dual theory shares this property of ``Legendre self-duality'', and we prove this here by using the CH formula \eqref{gensol} for the general self-dual NLED theory. A subsequent clarification of Kuzenko and Theisen was the observation that ``Legendre self-duality'' relies only on invariance under a discrete $Z_2$ subgroup of the $U(1)$ electromagnetic duality group \cite{Kuzenko:2000zw}. Here we give another proof based on the observation that if $\HH({\bf D},{\bf B})$ is invariant under ${\bf D}\leftrightarrow {\bf B}$ then a Legendre transform with respect to ${\bf D}$ must yield the same function as a Legendre transform with respect to ${\bf B}$. As an illustration, we explain how Born's original NLED theory of 1933 \cite{Born:1933lls} is Legendre self-dual without also being self-dual in the sense used here (and in \cite{Gaillard:1997rt,Kuzenko:2000zw}). 

We conclude, in a final section, with a summary of our main results and a brief discussion of further implications and future directions.

\section{Strong-field causality redux}\label{sec:redux}

As mentioned in the introduction, weak-field causality implies strong-field causality for self-dual NLED theories
if a weak-field limit is assumed. Without this assumption, causality requires the additional condition $G>0$, where
$G$ is given in \eqref{defG}. Either way, $G>0$ for causal theories and we can investigate its implications, as we did briefly in \cite{Russo:2024llm}. We now elaborate on some aspects
of this topic here because it will be useful when we later extend the results to the Hamiltonian formulation.

\begin{figure}[h!]
 \centering
 \begin{tabular}{cc}
\includegraphics[width=0.4\textwidth]{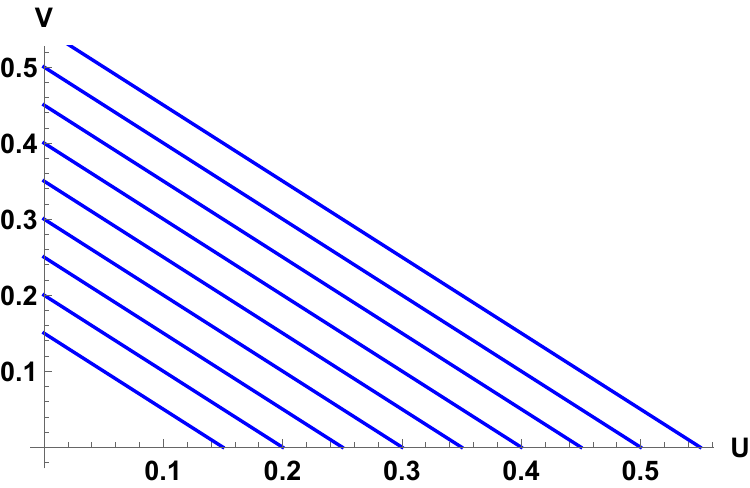}
 &
 \qquad \includegraphics[width=0.4\textwidth]{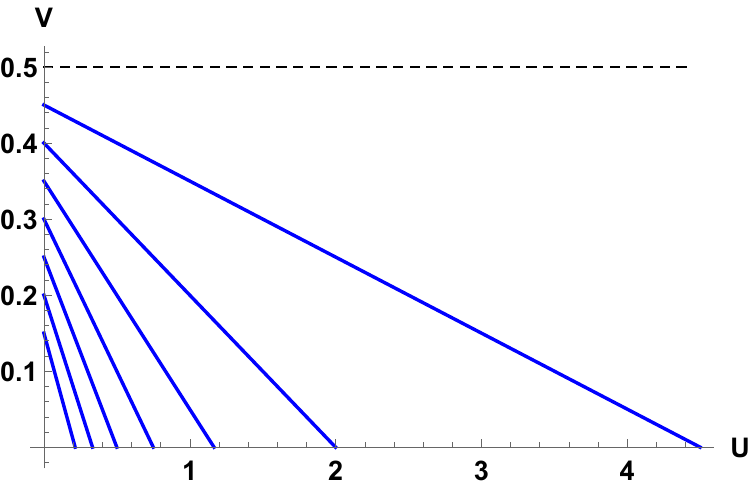}
 \\ (a)&(b)
 \end{tabular}
 \caption{The lines of constant $\tau$ for the two cases: (a) Maxwell, $\ell(\tau) =\tau$.
  (b) Born-Infeld, $\ell(\tau)=T-\sqrt{T(T-2\tau)}$
  (for $T=1$).
 }
 \label{fig:constantau}
 \end{figure}

From the equation for $\tau$ in \eqref{gensol} we learn that fixing $\tau$ restricts $(U,V)$ to a line in the positive-quadrant in the $(U,V)$ plane; i.e. the curves of constant $\tau$ in this quadrant are straight lines, with slopes 
\be
(dV/dU)(\tau) = - 1/[\dot\ell(\tau)|^2\, . 
\ee
Recalling the equation \eqref{intersect} for $d\tau$ we see that if $G=0$ at some point in the positive $(U,V)$ quadrant then we can take $\tau \to \tau+d\tau$ for $(U,V)$. In other words, 
$G=0$ at the intersection point of the lines of constant $\tau$ and constant $\tau+d\tau$. It follows that if $G>0$ everywhere in the domain of $\LL(U,V)$ (which is either the entire positive $(U,V)$ quadrant or a connected subregion of it that includes the origin) then no two lines of $\tau$ can intersect in this domain. This is because the line of constant $\tau$ can intersect the line of constant $\tau + c$, for any positive constant $c$, only if it also intersects the line of constant $\tau + d\tau$ for positive infinitesimal $d\tau$. Thus, if  $G>0$ in the domain of $\LL$ then this domain is foliated by lines of constant $\tau$, as illustrated for Maxwell and Born-Infeld in Fig.~\ref{fig:constantau}.

We can interpret this conclusion in another way. If the solution of \eqref{gensol} for $\LL(U,V)$
is unique then $\tau$ is uniquely determined by $(U,V)$ at each point in the domain 
of $\LL(U,V)$. However, there are at least two distinct values for $\tau$ at an intersection 
point, so $\tau$ cannot be uniquely determined by $(U,V)$ in any region that includes an intersection point. A necessary and sufficient condition for the uniqueness of the solution \eqref{gensol} is therefore that $G$ is nowhere zero in the domain of $\LL(U,V)$.
In those cases for which $\ell(\tau)$ has a power-series expansion of the form \eqref{series}, we know that $G>0$ is implied by the causality/convexity inequalities of \eqref{causal-L}, and hence that the solution for $\tau$ will be unique if these inequalities on $\dot\ell$ and $\ddot\ell$ are satisfied. Given the importance of this point, we shall show how it can be deduced in a more direct way. 

\begin{figure}[h!]
 \centering
 \begin{tabular}{cc}
\includegraphics[width=0.4\textwidth]{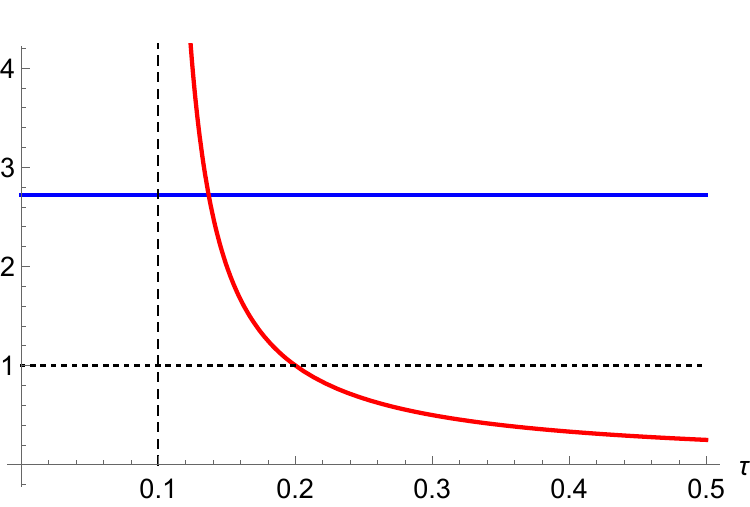}
 &
 \qquad \includegraphics[width=0.4\textwidth]{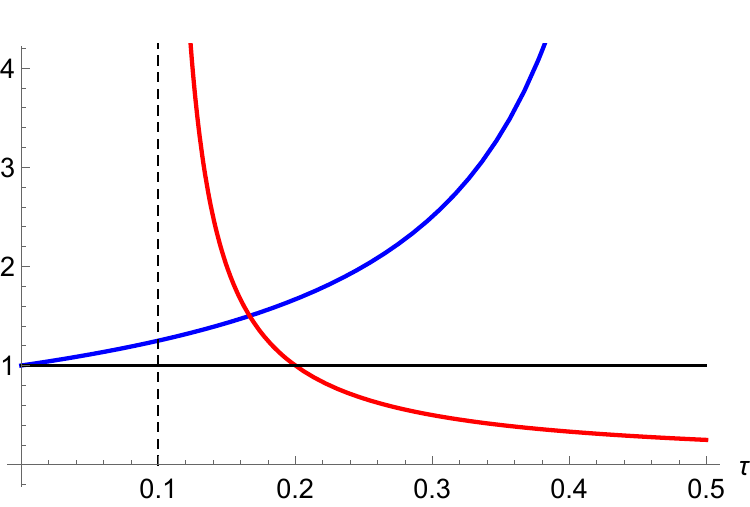}
 \\ (a)&(b)
 \end{tabular}
 \caption{Unique intersection (determining $\tau$)
 of the graph of $f(\tau)$ (blue curve) with the graph 
 of $F_{(U,V)}(\tau)$ (red curve) for a) ModMax ($\ell=e^\gamma \tau$ with $\gamma=0.5$),  and b) Born-Infeld ($\ell=T-\sqrt{T(T-2\tau)}$).
 }
 \label{fig:intertau}
 \end{figure}

We first rewrite the equation for $\tau$ in \eqref{gensol} as
\be\label{2nd}
f(\tau) = F_{(U,V)}(\tau) \, ; \qquad f := \dot\ell^2\, , \quad F_{(U,V)} := \frac{U}{\tau-V}\, . 
\ee
The solution for $\tau$ is unique if the graph of the function $f$ has precisely one intersection with the graph of the function $F_{(U,V)}$, for any choice of $(U,V)$ in the domain of $\LL$.  The function $f$ has the properties
\be
f(0)\geq 1\, , \qquad \dot f(\tau) \ge 0\, , 
\ee
which follow from \eqref{series} and \eqref{causal-L}; i.e. $f(\tau)$ is a function of non-negative slope for all $\tau>0$, with the minimum value being $f(0)\geq 1$.  The graph of the function $F_{(U,V)}(\tau)$ for $U>0$ is the branch of a hyperbola that has the line $\tau=V$ as one asymptote, and the $\tau$-axis as the other asymptote. From this description it is obvious that the graphs of the two functions intersect at precisely one point for each choice of $(U,V)$, 
which confirms that \eqref{2nd} has a unique solution for $\tau$, as illustrated in 
Fig.~\ref{fig:intertau} for ModMax and Born-Infeld. 

\subsection{Auxiliary fields and the stress-energy tensor}

It was observed in \cite{Russo:2024llm} that the two equations of \eqref{gensol} may be combined (for causal theories) into the single equation 
\be\label{formula}
\LL(U,V; \lambda,\tau) = \ell(\tau) - \frac{2U}{\dot\ell(\tau)} - \lambda \left(\tau - V - \frac{U}{[\dot\ell(\tau)]^2} \right)\, , 
\ee
where $\lambda$ is a Lagrange multiplier. This is because $\lambda$ and $\tau$ are, jointly,  a pair of auxiliary fields that can be consistently eliminated by their algebraic field equations. Varying 
$\tau$ yields the equation $G(\lambda- \dot\ell) =0$, which implies $\lambda=\dot\ell$ if $G>0$. Varying $\lambda$ yields a constraint that uniquely determines $\tau$ when $G>0$ (as illustrated in the previous subsection).  Elimination of $(\lambda,\tau)$ thus yields the Lagrangian density defined by \eqref{gensol}. Since $(U,V)$ are parity-even, we might expect to be able to make parity assignments for the auxiliary fields $(\lambda,\tau)$ such that the Lagrangian density of \eqref{formula} has even parity. 
This is true: if we assign even parity to both $\lambda$ and $\tau$ then both $\ell(\tau)$ and $\dot\ell(\tau)$ are parity-even, and hence so is $\LL(U,V;\lambda,\tau)$. 

An implicit assumption in the definitions of $(S,P)$, and hence of $(U,V)$, is that the Minkowski spacetime metric is the standard Minkowski metric (with a ``mostly plus'' signature). To generalize to curvilinear coordinates $\{x^\mu; \mu=0,1,2,3\}$, we have only to define $(S,P)$ as the scalar fields 
\begin{equation}
    S = -\frac14  \, {\rm g}^{\mu\rho} {\rm g}^{\nu\sigma}\, F_{\mu\nu}F_{\rho\sigma}\, ,  \qquad 
    P = -\frac{1}{8\sqrt{|{\rm g}|}}  \varepsilon^{\mu\nu\rho\sigma} F_{\mu\nu}F_{\rho\sigma}\, , 
\end{equation}
where ${\rm g}$ is the Minkowski metric in the chosen coordinates  
(with $|{\rm g}| = -\det{\rm g}$) and $F=dA$ is the 2-form abelian field strength for 1-form potential $A$ on the Minkowski spacetime.  It then follows that $(U,V)$ are scalars and hence so is $\tau$ and $\ell(\tau)$, and the equations of \eqref{gensol} still apply but with $\mathcal{L}$ a scalar rather than a scalar density. With this understood, \eqref{formula} is unchanged but the Lagrangian scalar density is now 
\be\label{Lgrav}
\mathfrak{L} :=    \sqrt{|{\rm g}|}\, \mathcal{L} =  \sqrt{|\rm g|}\,  \left[\ell (\tau) -\lambda\tau\right] + 
\left[\frac{\lambda-2\dot\ell}{\dot\ell^2}\right] \mathcal{U} + \lambda \mathcal{V}\, , 
\ee
where 
$(\mathcal{U}, \mathcal{V})$ are the scalar densities $\sqrt{|{\rm g}|}\, (U,V)$, which are related to the scalar densities 
$(\mathcal{S}, \mathcal{P}) =  \sqrt{|{\rm g}|}\, (S,P)$
in the same way that $(U,V)$ are related to $(S,P)$.

We are not restricted to Minkowski spacetime; by re-interpreting the metric ${\rm g}$ as an arbitrary spacetime metric that can be freely varied, we can find the stress-energy tensor $\TT$ by the Hilbert formula 
\be
\TT_{\mu\nu} = -\frac{2}{\sqrt{|{\rm g}|}} \frac{\partial \mathfrak{L}}{\partial {\rm g}^{\mu\nu}} \, . 
\ee
Since $\mathcal{P}$ is metric independent, this formula yields 
\be
\TT_{\mu\nu} = (\ell - \lambda\tau) {\rm g}_{\mu\nu}
+ \left\{ \lambda \frac{\partial\mathcal{V}}{\partial \mathcal{S}} + \left[ \frac{\lambda -2\dot\ell}{\dot\ell^2}\right]  
\frac{\partial\mathcal{U}}{\partial \mathcal{S}}\right\} \TT^{\rm Max}_{\mu\nu}\,  , 
\ee
where
\be
\TT^{\rm Max}_{\mu\nu} := -\frac{2}{\sqrt{|{\rm g}|}} \frac{\partial \mathcal{S}}{\partial {\rm g}^{\mu\nu}} \, , 
\ee
which is the Maxwell stress-energy tensor. Using 
\be
\frac{\partial \mathcal{V}}{\partial \mathcal{S}} = \frac{\partial V}{\partial S}=  \frac{V}{U+V} \, , \qquad 
\frac{\partial \mathcal{U}}{\partial \mathcal{S}} = \frac{\partial U}{\partial S}  =- \frac{U}{U+V} \, , 
\ee
and the auxiliary-field equations, we 
can simplify this result to\footnote{We thank Dmitri Sorokin for pointing out an error in the stress-energy tensor formula appearing in the original arXiv version of \cite{Russo:2024llm}.}
\be\label{SDTT}
\TT_{\mu\nu} = 
\left[\frac{\tau\dot\ell}{U+V}\right] \TT^{\rm Max}_{\mu\nu} + (\ell - \tau \dot\ell) {\rm g}_{\mu\nu}\, ,  \qquad \left(\tau=V+ \dot\ell^{-2}U\right). 
\ee
This agrees with the result of \cite{Russo:2024xnh}; the novelty here is that we have taken as our starting point the auxiliary-field formulation \eqref{Lgrav} for the Lagrangian density of a generic self-dual NLED in a general spacetime.

\section{The self-dual NLED Hamiltonian}\label{sec:SDH}

We have seen in the Introduction that the Hamiltonian density for the general self-dual and Lorentz invariant NLED
may be expressed in terms of a one-variable CH function $\hh(\sigma)$ via the equations of \eqref{generalh}. This 
was by analogy to the equations of \eqref{gensol} for the Lagrangian density, and the same steps may be used here to verify it.  By taking the exterior derivative of both 
sides of both equations of \eqref{generalh} we find two equations that are 
jointly equivalent to 
\be\label{ediff}
d\HH  = \hh^\prime dv + \frac {du}{\hh^\prime}\, , \qquad \tilde G d\sigma = \hh^\prime\left[(\hh^\prime)^2 dv - du\right] \,, 
\ee
where $\tilde G$ is the Hamiltonian analog of the function $G$ of \eqref{defG}:
\be\label{tildeG}
\tilde G =  (\hh^\prime)^3 - 2u\hh^\prime{}^\prime\, . 
\ee
We shall see later that causality requires $\tilde G>0$, with consequences analogous to those that follow from $G>0$.

The first equation of \eqref{ediff} tells us that
\be\label{Huv}
\HH_v = \hh^\prime\, , \qquad \HH_u = 1/\hh^\prime \,, 
\ee
and hence that $\HH$ solves \eqref{GR-H}.  Notice that any constant term in $\hh(\sigma)$, which makes no contribution to 
$\hh^\prime(\sigma)$, appears only as a constant term in $\HH$; it represents a constant uniform background energy 
density that has no effect on the NLED field equations. 

Our first task will be to determine the relation between the functions $\ell$ and $\hh$ implied by Legendre duality
of the Lagrangian and Hamiltonian densities. The existence of this duality is guaranteed by the convexity
of $\LL$ as a function of ${\bf E}$, the fact that the Legendre transform of any function is convex, and the theorem that 
the Legendre transform is an involution when acting on convex functions. For ${\bf B}={\bf 0}$, this transform is 
 \be\label{LTzeroB}
 \begin{aligned}
   \LL({\bf E},{\bf 0}) &= \sup_{\bf D} \left\{{\bf D}\cdot {\bf E}-\HH({\bf D},{\bf 0})\right\}\, ,     \\
  \HH({\bf D}, {\bf 0}) &= \sup_{\bf E} \left\{{\bf E}\cdot {\bf D}-\LL({\bf E},{\bf 0})\right\}\, .  \\
\end{aligned}
 \ee
 When ${\bf B}={\bf 0}$ we also have
\be
\begin{aligned}
(U,V) &= (0,\tau)\, , \qquad \tau = \frac12 E^2\, , \\
(u,v) &= (0,\sigma)\, , \qquad \sigma= \frac12 D^2   \, , 
\end{aligned}
\ee
and hence, from \eqref{gensol}  and \eqref{generalh}, 
\be
\begin{aligned}
\LL({\bf E},{\bf 0})  &= \ell(\tau) = L(E) \, , \\
 \HH({\bf D}, {\bf 0}) &= \hh(\sigma) = H(D)\, , 
\end{aligned}
\ee
where $L$ and $H$ are the functions introduced in \eqref{L&H}.  Combining this with \eqref{LTzeroB}, we have
 \be\label{LH1}
  \begin{aligned}
  L(E) &= \sup_{\bf D} \left\{{\bf D}\cdot {\bf E}- H(D)\right\}\, ,     \\
  H(D) &= \sup_{\bf E} \left\{{\bf E}\cdot {\bf D}-L(E)\right\}\, .  \\
\end{aligned}
 \ee
 
 Notice that although $L(E)$ and $H(D)$ were defined in \eqref{L&H} as functions of a single variable
 (respectively, $E=\sqrt{2\tau}$ and $D=\sqrt{2\sigma}$), we are required by \eqref{LH1}  to consider them 
 as functions of ${\bf E}$ and ${\bf D}$, respectively. In contrast, the  claim in the Introduction that 
 $L$ and $H$ are each other's Legendre transform is the claim that 
 \be\label{LH2}
  \begin{aligned}
  L(E) &= \sup_D \left\{DE- H(D)\right\}\, ,     \\
  H(D) &= \sup_E \left\{ED-L(E)\right\}\, .   
\end{aligned}
 \ee
 However, it is not difficult to see that \eqref{LH1} implies \eqref{LH2}. 
Variation of ${\bf D}$ and ${\bf E}$ in the respective expressions 
 of \eqref{LH1} for $L(E)$ and $H(D)$ yields
 \be
 \begin{aligned}
 {\bf E} &= \left(\frac{1}{D} \frac{\partial H}{\partial D}\right) {\bf D} = \hh^\prime(\sigma) {\bf D} \quad 
 \Rightarrow \quad {\bf D}\cdot{\bf E} = DE\, ,  \\
{\bf D} &= \left(\frac{1}{E} \frac{\partial L}{\partial E}\right) {\bf E} = \dot\ell(\tau) {\bf E}  \qquad 
\Rightarrow\quad {\bf E}\cdot{\bf D} = ED\, , 
 \end{aligned}
 \ee
 and a further implication is 
\be\label{varLH2}
E = \hh^\prime (\sigma) D \, , \qquad D = \dot\ell(\tau) E \, , 
\ee
which is exactly what one finds from variation of  $D$ and $E$ in the  expressions of \eqref{LH2} for $L(E)$ and $H(D)$, respectively.
The variation of 3-vector fields needed to find the functions $L$ and $H$ from \eqref{LH1} therefore yields the same result
as the variation of scalar fields in \eqref{LH2}. 

Further implications of \eqref{varLH2} are the relations\footnote{A relation similar to \eqref{dlph} appears in \cite{Perry:1996mk} in relation to an involution defined in the context of 6D chiral electrodynamics.}
\be\label{dlph}
\dot\ell(\tau) \hh^\prime(\sigma)  =1\, , 
\ee
and 
\be\label{sigtau}
\sigma = \tau \dot\ell^2 \, , \qquad \tau = \sigma (\hh^\prime)^2\, .
\ee
These relations allow us to find $\hh(\sigma)$ (up to the addition of a constant) given $\dot\ell(\tau)$, and {\it vice versa}. 

The fact that functions $\frak{h}$ and $\ell$ 
are related by  Legendre transformations, but with respect to  variable
$\sqrt{2\tau}$ and $\sqrt{2\sigma}$, can be summarized by  the equations
\be\label{ell+h}
\ell(\tau) + \frak{h}(\sigma) = 2\tau \dot\ell =  2 \sigma \frak{h}^\prime\, . 
\ee
The second equality is equivalent to \eqref{sigtau} given \eqref{dlph}. The first equality tells us that any constant term 
in the power-series expansion of $\ell(\tau)$ also appears in the power-series expansion of $\frak{h}(\sigma)$ with the 
opposite sign, while the remaining information of this equality may be verified by taking the exterior derivative of both sides
to get 
\be
\frak{h}^\prime d\sigma = (\dot \ell + 2\tau \ddot\ell )d\tau\, . 
\ee
Using \eqref{dlph}, we may rewrite this as $d\sigma= d\left(\tau\dot\ell^2\right)$, which is 
true as a consequence of the relation $\sigma= \tau\dot\ell^2$ of \eqref{sigtau}. 

There are other differences between the Lagrangian and Hamiltonian formulations 
of self-dual NLED theories that go beyond sign changes. One is the difference in the potential range of the independent variables: although $\tau$ is non-negative by its definition in \eqref{gensol}, the definition of $\sigma$ in \eqref{generalh}, which we may rewrite as 
\be\label{negsig}
\sigma = (v-u) + \frac{ [(\hh^\prime)^2 -1]u}{(\hh^\prime)^2}\, ,  
\ee
allows $\sigma$ to be negative. For $\hh^\prime=1$, which is the free-field (Maxwell) case, 
$\sigma= v-u\ge0$, and this remains true for all causal  NLED theories that have Maxwell as their weak-field limit. This is easily seen by writing the second equation in \eqref{generalh} as
\be\label{fandg}
f(\sigma):= (\hh^\prime)^2(\sigma)= \frac{u}{v-\sigma} := g(\sigma)\, . 
\ee
The function $f(\sigma)$ is positive, with $f(0)=1$ (because the weak-field limit is Maxwell). It is also a monotonically decreasing function of $\sigma$ (because $\hh'>0$ but $\hh''<0$ for any causal interacting NLED). The function $g(\sigma)$ takes the value $u/v \le 1$ at $\sigma=0$ but then increases monotonically, becoming infinite at $\sigma=v$, and then negative. There is therefore a unique non-negative value of $\sigma$ at which $f=g$ (as illustrated in fig. \ref{fig:intersigma}a below). 

In contrast, if the weak-field limit is ModMax (with $\gamma>0$) then $f(0)<1$. This means that there will be a choice of $(u,v)$ such that $f(0)<g(0)$, which implies that $f=g$ for $\sigma<0$, as illustrated in fig. \ref{fig:intersigma}b. In these cases $\sigma<0$ is not excluded by its definition in \eqref{generalh}; instead the inequality $\sigma\ge0$ is a restriction  on the domain of $\HH$ required by equivalence to the Lagrangian formulation. Specifically, it restricts the Hamiltonian fields to the region in field-space for which $\HH(u,v)$ is a convex function of ${\bf D}$; i.e. to its ``convex domain''. For ModMax, the boundary of this convex domain corresponds to Lagrangian fields with $U=V=0$, which includes all exact plane-wave solutions of the ModMax field equations\cite{Bandos:2020jsw}.

\subsection{Convexity/Concavity and Causality}

In the Lagrangian formalism, and assuming the existence of a weak-field limit, the necessary and sufficient conditions for 
causality are the conditions for convexity of $\mathcal{L}$, which are equivalent to \cite{Russo:2024llm}
\be\label{ccc}
\dot \ell(\tau ) \geq 1\ ,\qquad \ddot \ell \ge 0\, .  
\ee
By combining the first of these inequalities with the relation \eqref{dlph} we deduce that 
\be\label{hprime-ineq}
0< \hh^\prime \le1\, . 
\ee
Next, we take the exterior derivative of the first of the relations in  \eqref{dlph} to find, again using \eqref{dlph}, that
\be\label{hpp}
\dot\ell^2 \hh^\prime{}^\prime =  -\left(\frac{d\tau}{d\sigma}\right)\ddot\ell \, . 
\ee
Taking the  exterior derivative of the first of the relations of \eqref{sigtau}, we also find that 
\be\label{dtds}
\frac{d\tau}{d\sigma} = \frac{1}{\dot\ell\left(\dot\ell + 2\tau \ddot\ell\right)}\, ,  
\ee
and hence that 
\be
-\hh^\prime{}^\prime =  \frac{\ddot \ell}{\dot\ell^3\left(\dot\ell + 2\tau \ddot\ell\right)}\, . 
\ee
Using both inequalities of \eqref{ccc}, and the fact that $\tau$ is non-negative, we see that 
the right-hand side of this equation is non-negative, and hence 
\be
\hh^\prime{}^\prime \le 0\, . 
\ee
We have now found the `translation' of the causality/convexity conditions  \eqref{ccc} on $\ell(\tau)$  to the corresponding 
conditions to be satisfied by $\hh(\sigma)$. They are 
\be\label{ccch}
0< \hh^\prime \le1\, , \qquad \hh^\prime{}^\prime \le 0\, . 
\ee
The second of these equations is equivalent to the statement that  $\hh(\sigma)$ is a concave function, 
and a corollary of this is that 
\be\label{tildeG0}
\tilde G >0\, , 
\ee
where $\tilde G$ was defined in \eqref{tildeG}. We postpone a discussion of the consequences of this corollary 
as we still need to explain the origin of the condition \eqref{hprime'} of the Introduction.

By taking the exterior derivative on both sides of the second of the relations of \eqref{sigtau}, 
we get another formula for $d\tau/d\sigma$:
\be
\frac{d\tau}{d\sigma} = \hh^\prime (\hh^\prime + 2\sigma \hh^\prime{}^\prime) \, . 
\ee
Comparing this with \eqref{dtds}, and again using \eqref{dlph}, we find that
\be\label{iden}
\big(\dot\ell + 2\tau\ddot\ell\big) \big(\hh^\prime + 2\sigma \hh^\prime{}^\prime\big) =1\, . 
\ee
The first factor on the left-hand side is positive, for reasons explained above. The second factor is not obviously positive, 
but is required to be so; this is the condition \eqref{hprime'}. To understand its significance, we return to  the functions $L(E)$ and $H(D)$. Because they are each other's Legendre transform, we know that they are both convex; in fact strictly convex because $\mathcal{L}$ is a strictly convex function of ${\bf E}$. Thus 
\be
0  < \frac{\partial^2 L(E)}{\partial E\partial E} = \dot\ell + 2\tau \ddot\ell\, , \qquad 
0  < \frac{\partial^2 H(D)}{\partial D\partial D} = \hh^\prime + 2\sigma \hh^\prime{}^\prime\, . 
\ee
This allows us to interpret \eqref{iden} as the statement that the Hessian of $L(E)$ is the inverse of the Hessian of 
$H(D)$. This requires, of course, that both Hessians are non-zero and finite, which is equivalent to the statement 
that both $L(E)$ and $H(D)$ are both {\sl strictly} convex functions.

We now return to the significance of \eqref{tildeG0}. We see from \eqref{generalh} that the curves of constant $\sigma$ in the $(u,v)$-plane are straight lines. Only the 
half-lines in the `physical' region of this plane are relevant; this is the wedge-shaped region bounded by the lines $u=0$ (the $v$-axis) and the line $v=u$ (since $v\ge u\ge0$ by definition). Because $\tilde G>0$, no two lines of constant $\sigma$ can intersect in this region (for reasons identical to those explained in our discussion of 
section \ref{sec:redux} for $G>0$ in the context of lines of constant $\tau$ in the positive 
$(U,V)$ quadrant). The lines of constant $\sigma$ therefore foliate either the entire physical region or some connected subregion of it. 

\begin{figure}[h!]
 \centering
 \begin{tabular}{ccc}
\includegraphics[width=0.3\textwidth]{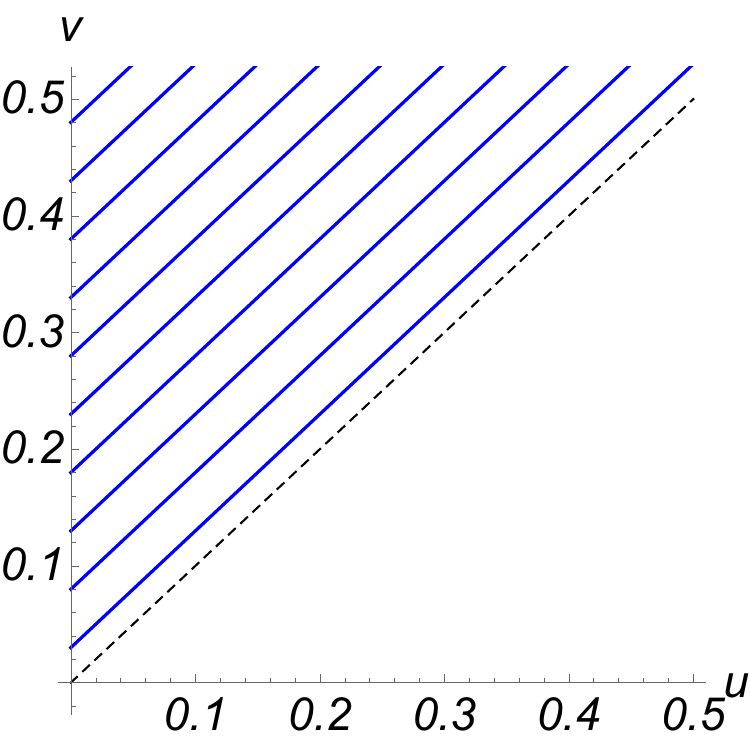}
 &
 \ \  \includegraphics[width=0.3\textwidth]{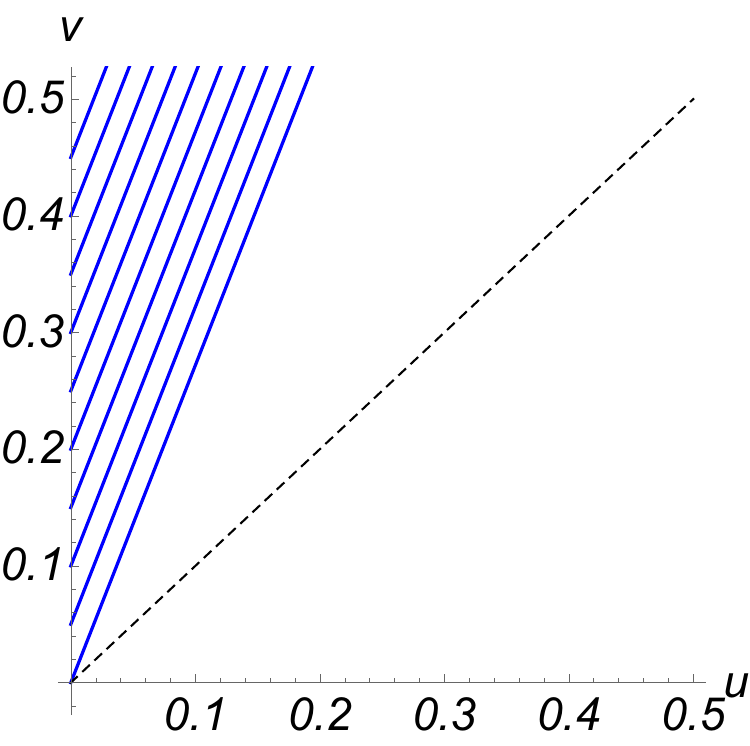}
 &
 \ \  \includegraphics[width=0.3\textwidth]{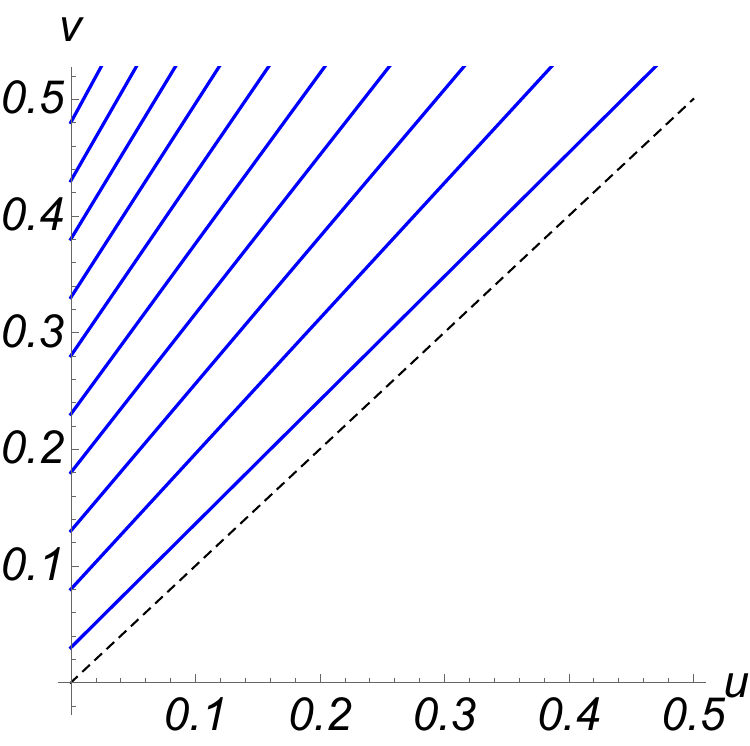}
 \\ (a)&(b)&(c)
 \end{tabular}
 \caption{Lines of constant $\sigma$, defined in \eqref{generalh}, for (a) Maxwell, $h(\sigma) =\sigma$,  (b) ModMax, $h(\sigma) =e^{-\gamma}\sigma$,  $\gamma =0.5$. and (c) Born-Infeld, $h(\sigma)=\sqrt{T(T+2\sigma)}$ (for $T=1$).
 }
 \label{fig:constantsigma}
 \end{figure}

From \eqref{generalh} we see that all  lines of constant $\sigma$ intersect the $v$-axis at $v=\sigma$, which implies (since $v\ge0$) that the lowest line is the one with $\sigma=0$; this confirms that $\sigma\ge0$ is required for an equivalence of the Lagrangian and Hamiltonian formulations. The slope of the lines is 
\be
(dv/du)(\sigma)= 1/[\hh^\prime(\sigma)]^2\ge 1 \, , 
\ee
where the inequality follows from the first causality condition of \eqref{ccch}. The slope of the lowest line is therefore 
$1/[\hh^\prime(0)]^2$. Assuming that $\hh(\sigma)$ has a power-series expansion about $\sigma=0$ (which is equivalent to the assumption of a weak-field limit) we conclude (omitting the irrelevant constant term in the expansion) that 
\be
\hh(\sigma) = e^{-\gamma}\sigma + \mathcal{O}(\sigma^2)\, , 
\ee
for some constant $\gamma\ge0$. The special case for which $\hh(\sigma)= e^{-\gamma}\sigma$ yields ModMax, with Maxwell as the free-field $\gamma=0$ subcase. For Maxwell, the lines of constant $\sigma$ foliate the entire wedge-shaped physical region in the $(u,v)$-plane. For ModMax ($\gamma>0$) they foliate the wedge-shaped subregion that is bounded from below by the $\sigma=0$ line, which is 
\be
v= e^{2\gamma} u  \qquad (\sigma=0).
\ee
For both Maxwell and ModMax the lines of constant $\sigma$ are parallel because $h'$ is constant. For BI, the slope increases as $\sigma$ increases because $\hh^\prime{}^\prime <0$. These three cases are illustrated in Fig.~\ref{fig:constantsigma}.

For the examples that we consider in the following section, there is no maximum value of $\sigma$, so the lines of constant $\sigma$ foliate the wedge-shaped region bounded by the positive $v$-axis and the $\sigma=0$ line. We have found an example with an upper bound on $\sigma $ but we do not discuss it here. 

\begin{figure}[h!]
 \centering
 \begin{tabular}{cc}
\includegraphics[width=0.4\textwidth]{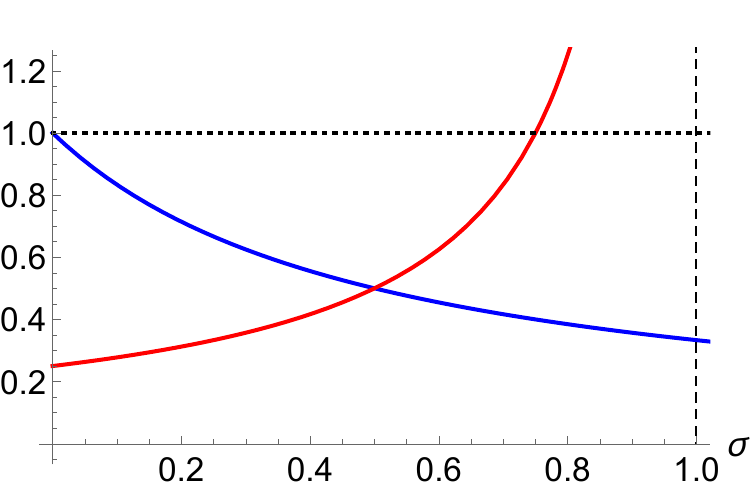}
 &
 \qquad \includegraphics[width=0.4\textwidth]{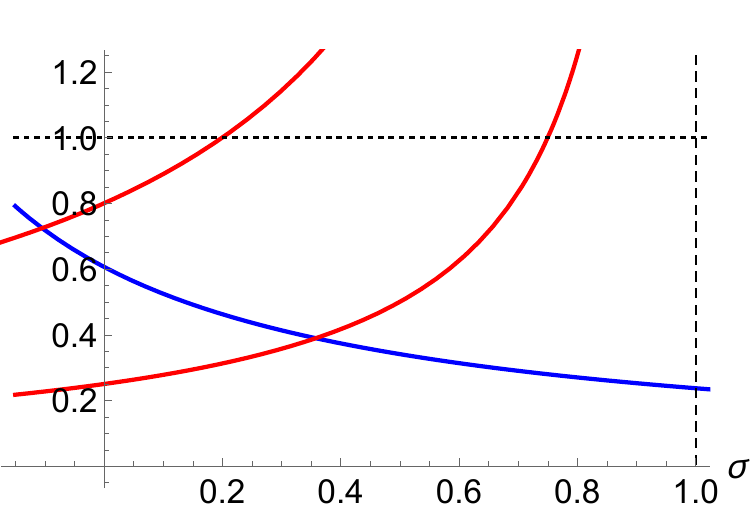}
 \\ (a)&(b)
 \end{tabular}
 \caption{Unique intersection (determining $\sigma$) of the graph of $ (\hh'(\sigma))^2$ (blue curve) with the graph of $u/(v-\sigma)$ (red curve) for: a) Born-Infeld at $u=v/4$. For any $u,v$, $u\leq v$, intersection occurs at $\sigma\geq 0$. b)   ModMaxBorn with $\gamma=0.5$. When $u=v/4$ the intersection occurs at $\sigma\geq 0$, but when $u>ve^{-2\gamma}$ the intersection occurs at $\sigma<0$  ($T=1$).
 }
 \label{fig:intersigma}
 \end{figure}

A further implication of $\tilde G>0$ is that there is a Hamiltonian counterpart to \eqref{formula}. The two equations of \eqref{generalh} may be combined into the one equation 
\be\label{formulaH}
\HH = \hh(\sigma)+\frac{2u}{\hh'(\sigma)} - \tilde\lambda \left(\sigma- v+\frac{u}{\left[\hh'(\sigma)\right]^2}\right)\, , 
\ee
where $\tilde\lambda$ is a Lagrange multiplier imposing the constraint on $\sigma$, but the fields $(\tilde\lambda, \sigma)$ are an auxiliary pair. Varying $\sigma$ we get the equation 
$\tilde G(\tilde\lambda - \hh^\prime)=0$, which is equivalent to $\tilde\lambda = \hh^\prime$ when
$\tilde G>0$. Varying $\tilde\lambda$ we get the equation for $\sigma$, which has a unique solution when $\tilde G>0$ for reasons identical to those
explained in section \ref{sec:redux} for $G>0$ in the context of the equation for $\tau$. This is illustrated in Fig.~\ref{fig:intersigma}. 
Elimination of the auxiliary fields in \eqref{formulaH} therefore yields precisely the Hamiltonian density 
defined by \eqref{generalh}.  

As for the Lagrangian auxiliary-field formulation of \eqref{formula},
we can make parity assignments for the Hamiltonian auxiliary fields $(\tilde\lambda,\sigma)$ such that the Hamiltonian density of \eqref{formulaH} has even parity. Since both $u$ and $v$ are parity-even, this is achieved by assigning even parity to both $\tilde\lambda$ and $\sigma$.  A consequence of parity conservation is that the $U(1) \cong SO(2)$ electromagnetic duality group is enhanced 
to $O(2)$ because parity acts by the transformation ${\bf D} \to - {\bf D}$.

\subsection{Simple examples}\label{sec:examples}

We now illustrate the construction of the  Hamiltonian
from $\hh $ and the causality conditions on $\hh $ with a few examples.

\subsubsection*{ModMax}

The ModMax Lagrangian and Hamiltonian densities are\footnote{Recall that 
$(u,v)$ as defined in this paper differ from the definitions in \cite{Bandos:2020jsw}  
by $u\leftrightarrow v$.} \cite{Bandos:2020jsw} 
\be
\begin{aligned}
\LL_{MM} &= e^\gamma V -e^{-\gamma} U = (\cosh\gamma)S + (\sinh\gamma)\sqrt{S^2+P^2}\, , \\
\HH_{MM} &= e^{-\gamma }v+ e^{\gamma } u = (\cosh\gamma) s - (\sinh\gamma) \sqrt{s^2-p^2}\, .  
\end{aligned}
\ee
Maxwell is included as the special case with $\gamma=0$.  The Lagrangian CH-function for ModMax
is $\ell =e^\gamma \tau + {\rm const.}$ \cite{Russo:2024llm} (but we may ignore the constant term as it has 
no effect on the field equations). The convexity/causality conditions  of \eqref{causal-L} require $\gamma\ge0$, as 
expected since $\LL_{MM}$ is a convex function of ${\bf E}$ for $\gamma\ge0$ but not for $\gamma<0$.   
The function $\ell(\tau)=e^\gamma \tau$  corresponds to $L(E)= \frac12e^\gamma E^2$. Its Legendre transform is
 $H(D)= \frac12 e^{-\gamma}D^2$, which yields 
\be
\hh(\sigma) = e^{-\gamma } \sigma\, .  
\ee
Using this in \eqref{generalh} we have 
\be
\HH =e^{-\gamma } \sigma+2u e^{\gamma }\ ,\qquad \sigma =v- e^{2\gamma}u \ , 
\ee
which gives us the ModMax Hamiltonian density. 

As shown in \cite{Bandos:2020jsw}, $\HH_{MM}$ is a convex function of ${\bf D}$ for $\gamma>0$ only for those values of $({\bf D},{\bf B})$ for which 
\be\label{bound}
s \ge (\cosh\gamma) p\, .
\ee
Values of $({\bf D},{\bf B})$ violating this bound do not correspond to any values of $({\bf E},{\bf B})$. 
In other words, the bound is needed for a correspondence between the Lagrangian and Hamiltonian 
formulations of ModMax. However, we have seen in section \ref{sec:SDH}, for {\sl any} self-dual NLED, that 
this correspondence exists iff $\sigma\ge0$. It follows that the ModMax convexity bound \eqref{bound} 
must be equivalent to $\sigma\ge0$, and this conclusion is easily verified: from \eqref{generalh} we see that\
\be
\sigma\ge0 \quad \Leftrightarrow \quad v \ge e^{2\gamma} u \, , 
\ee
but this constraint on the values of $(u,v)$ is equivalent to \eqref{bound}.

\subsubsection*{ModMaxBorn}

The Born-Infeld-type generalization of ModMax, introduced in \cite{Bandos:2020jsw} in its Hamiltonian formulation,  was called
ModMaxBorn in \cite{Russo:2024kto,Russo:2024llm}. The ModMaxBorn Lagrangian density was found by Legendre transform 
in \cite{Bandos:2020hgy}. The Lagrangian and Hamiltonian densities are, respectively, 
\be
\begin{aligned}
\LL_{\rm MMB} &= - \sqrt{T^2 -2T\LL_{MM}- P^2} \, , \\
\HH_{\rm MMB} &= \sqrt{T^2 +2T\HH_{MM} + p^2}\, . 
\end{aligned}
\ee
The Born-Infeld theory is included as the $\gamma=0$ case.  Here and in other  examples to follow, there is a non-zero vacuum energy, which  can be simply removed by the addition of a constant.

 It was shown in \cite{Russo:2024llm} that $\LL_{\rm MMB}$ is found from the CH-function 
\be
\ell_{\rm MMB}(\tau) = - \sqrt{T(T- 2 e^{\gamma} \tau)}=-T \left(1- \frac{2 e^{\gamma} \tau}{T}\right)^{\frac12}\, . 
\ee
From the results of section \ref{sec:SDH} we find that the Hamiltonian CH-function $\hh(\sigma)$ is
\be
\hh_{\rm MMB}(\sigma) =\sqrt{T(T+ 2 e^{-\gamma} \sigma)}=T \left(1+ \frac{2 e^{-\gamma} \sigma}{T}\right)^{\frac12}\, ,   
\ee
Using this result in \eqref{generalh} we recover the ModMaxBorn Hamiltonian density,  {\sl and} we find that 
\be
\sigma = \frac{T(v-e^{2\gamma}u)}{T+ 2 e^\gamma u}\, .  
\ee
For $\gamma>0$ this allows $\sigma<0$, but for reasons already explained
we must impose $\sigma\ge0$, which is again equivalent to the bound \eqref{bound}. 

\subsubsection*{$q$-deformed $\hh _{\rm MMB} $}

Consider the following choice: 
\be
\label{hhqqq}
\hh =T \left(1+\frac{e^{-\gamma }\sigma}{qT}\right)^q\, ,   
\ee
for which 
\be
\hh' =e^{-\gamma}\left(1+\frac{e^{-\gamma }\sigma}{qT}\right)^{q-1} \, , \qquad 
\hh'' = -\frac{e^{-2 \gamma } (1-q)}{q T} 
   \left(1+\frac{e^{-\gamma } \sigma}{q
   T}\right)^{q-2}\, .  
\ee
From this we see that the conditions $\hh^\prime\leq 1$ and $\hh^\prime{}^\prime\le0$ require $0<q\leq 1$. 
We also have 
\be
\hh'+2\sigma \hh'' =e^{- \gamma }
   \left(1+\frac{e^{-\gamma } \sigma}{q
   T}\right)^{q-2} \left( 1+ (2 q-1)
  \frac{ \sigma e^{- \gamma }}{q T}\right)\, , 
\ee
which is positive, as required, if $q\geq \frac12$. Therefore,  this class of self-dual NLED theories is causal for 
\be
\frac12 \leq q \leq 1\, . 
\ee
However, the Hamiltonian density can be found explicitly only for special values of $q$ in this range; for example $q=\frac12$, which yields Born-Infeld. 
Another special choice is $q=\tfrac34$, which will be discussed later. 

\subsection{Conformal Invariance Redux}

The condition for conformal invariance of any Hamiltonian density $\mathcal{H}({\bf D},{\bf B})$ is degree-2 homogeneity in the electric and magnetic  fields $({\bf D},{\bf B})$. For a  self-dual NLED with Hamiltonian density function $\HH(u,v)$ this condition becomes degree-1 homogeneity in $(u,v)$:
\be\label{homog}
u\HH_u + v\HH_v = \HH\, .  
\ee
We recall here that our $(u,v)$ variables, defined in \eqref{uandv}, differ (by the exchange $u\leftrightarrow v$)
from those used in \cite{Bandos:2020jsw}.  The general solution of this equation can be expressed in the form
\be\label{conf}
\HH = v f(x)\,  , \qquad x:= u/v\, .
\ee
Notice that $u/v$ remains finite as $v\to 0$ since $u\le v$. The Lorentz invariance condition \eqref{GR-H} then implies that  $f'(f-xf') =1$. This equation is solved by (i) any linear function of $x$ and (ii) 
$f=\pm \sqrt{4x}$, which yield the following solutions for $\HH$ \cite{Bandos:2020jsw}:
\be
(i): \quad \HH = \tilde av + \tilde a^{-1} u\, , \qquad (ii): \quad \HH = \pm \sqrt{4uv}\, . 
\ee
The first of these is ModMax if $\tilde a= e^{-\gamma}$ with $\gamma\ge0$. The second solution defines (for positive sign)
the Bialynicki-Birula (BB) electrodynamics theory \cite{Bialynicki-Birula:1984daz,Bialynicki-Birula:1992rcm}, which has no weak-field limit. ModMax is therefore the unique  interacting causal `extension' of Maxwell electrodynamics with the same symmetries \cite{Bandos:2020jsw}. 

The condition for conformal invariance of any Lagrangian density $\mathcal{L}(S,P)$ is the homogeneity condition
\be\label{homog2}
S\mathcal{L}_S + P\mathcal{L}_P = \mathcal{L}\, . 
\ee
Any function linear in $(S,P)$ will satisfy this relation, but this does not include ModMax.  If parity is assumed then, as observed in the Introduction, one may replace the variables $(S,P)$ by $(S,\Phi)$ (we recall that $\Phi=\sqrt{S^2+P^2}$). The homogeneity condition is then solved by any function linear in $S$ and $\Phi$, and self-duality selects the
particular linear function that is the ModMax 
Lagrangian density, found originally by Legendre transform of the
ModMax Hamiltonian density \cite{Bandos:2020jsw}. This observation was made in \cite{Kosyakov:2020wxv}, but it does not exclude the possibility of other conformal self-dual NLED theories for which $\LL$ is a nonlinear homogeneous function of $(S,\Phi)$; for this we need a {\sl general} solution to the homogeneity condition \eqref{homog2}. 

One might expect to be able to express the general solution to \eqref{homog2} in terms of an arbitrary function $f$ of one dimensionless ratio of functions of $(S,P)$, by analogy to the general solution of \eqref{conf} to the Hamiltonian homogeneity condition \eqref{homog}. However, the fact that both $S$ and $P$ may have either sign, and may be zero for non-vacuum field configurations,  prevents it. For example,  the formula $\LL= \sqrt{SP} f(\sqrt{S/P})$ was suggested in \cite{Kosyakov:2007qc} but even $\LL=S$ cannot be written  in this form when $S<0$. The alternative formula $\LL = S f(P/S)$, suggested in \cite{Bandos:2020jsw}, has a similar problem with $\LL= \sqrt{S^2+P^2}$. If parity invariance is assumed then we may use the variables $(U,V)$, in which case \eqref{homog2} is replaced by 
\be\label{homog3}
V\mathcal{L}_V + U\mathcal{L}_U = \mathcal{L}\,  
\ee
In this case we could attempt to solve the homogeneity condition by 
setting $\LL(U,V) = Vf(U/V)$. This is the natural Lagrangian analog of \eqref{conf}, 
and imposing the self-duality condition leads formally to $f'(f-xf')=-1$; the different sign on the right-hand side now allows only a linear function of $x$, which again leads uniquely to ModMax. However this is still unsatisfactory because $U/V$ is generically infinite at $V=0$, so the initial expression for $\LL$ is not 
well-defined for all $(U,V)$.  

It appears that the only way to establish directly that the ModMax Lagrangian density is the 
unique possibility compatible with conformal invariance and self-duality is to first solve the self-duality condition, e.g. as in \eqref{gensol}. We then impose the homogeneity condition \eqref{homog3}. Using \eqref{partials1}, this leads to 
\be
V\dot\ell - \frac{U}{\dot\ell} = \ell -\frac{2U}{\dot\ell}
\ee
and hence 
\be
\ell = \left( V+ \frac{U}{\dot\ell^2} \right)\dot\ell  = \tau\dot\ell
\ee
where the last equality uses the definition of $\tau$ in \eqref{gensol}. We thus arrive at the conclusion, for self-dual NLED, that $\LL$ will satisfy the homogeneity condition 
\eqref{homog3} iff $\ell$ satisfies the homogeneity condition 
\be\label{homogl}
\tau\dot\ell(\tau) = \ell(\tau)\, . 
\ee
The general solution is $\ell(\tau)= a\tau$ for constant $a$. Causality restricts to 
$a\ge 1$, which yields ModMax.

\section{Hamiltonian without Legendre transform}\label{subsec:LDWL}

The solution \eqref{gensol} to the self-duality PDE \eqref{GR} results from a choice of boundary conditions on the $U=0$ boundary of the positive $(U,V)$ quadrant:  $\LL(0,V) = \ell(V)$.  However, we could equally well choose initial conditions on the $V=0$ boundary of the positive $(U,V)$ quadrant; i.e. $\LL(U,0) = -m(U)$ for some new one-variable function $m$ (the minus sign is included for later convenience). The solution analogous to \eqref{gensol} is then 
\be\label{gensol2}
\LL(U,V) = -m(\kappa) + \frac{2V}{\dot m(\kappa)} \, , \qquad \kappa  = U + \frac{V}{\dot m^2(\kappa)}\, ,  
\ee
where
\be
\dot m(\kappa)  := \frac{dm(\kappa)}{d\kappa}\,  >\, 0\, . 
\ee 
For the identity function $m(\kappa)=\kappa$ these equations yield $\LL= V-U =S$. To verify that they yield a solution for arbitrary $m(\kappa)$, we proceed as before by taking the differential of both sides of both equations to find that 
\be
d\LL = \frac{1}{\dot m} dV - \dot m dU\, , \qquad (\dot m^3 +2\ddot m V) d\kappa = \dot m\left(dV + \dot m^2 dU\right) .
\ee
From the first of these equations we have
\be\label{partials2}
\LL_U = - \dot m\, , \qquad \LL_V = 1/\dot m \, , 
\ee
and hence $\LL_U\LL_V =-1$, as required.   

We now have two different ways in which the Lagrangian density function $\LL(U,V)$ of any given self-dual NLED theory can be constructed from an associated one-variable function; in one case we call the function $\ell(\tau)$ and in the other case we call it $m(\kappa)$. By comparing \eqref{partials2} with \eqref{partials1} we see that these
two functions are such that\footnote{Recall that $\dot \ell=d\ell/d\tau$ and $\dot m=dm/d\kappa$.}
\be\label{ell-m}
\dot\ell(\tau) \dot m(\kappa) =1\, . 
\ee 
Using this relation, a comparison of the equation \eqref{gensol} for $\tau$ with equation \eqref{gensol2} for $\kappa$ provides 
an equation for $\tau$ as a function of $\kappa$, and {\it vice versa}:
\be\label{kaptau}
\tau= \kappa \dot m^2(\kappa) \, , \qquad \kappa= \tau\dot\ell^2(\tau)\, .
\ee 
If we use the relations \eqref{ell-m} and \eqref{kaptau} in the equation of \eqref{gensol2} for $\kappa$ we deduce that 
\be
\tau = V + \frac{U}{\dot\ell^2}\, , 
\ee 
which is the equation for $\tau$ of \eqref{gensol}.  Since the equations for the auxiliary variable ($\tau$ or $\kappa$) are equivalent in both solutions of the self-duality PDE, which yield the same Lagrangian density function, it follows that 
\be 
\ell(\tau) - \frac{2U}{\dot\ell(\tau)} = -m(\kappa) + \frac{2V}{\dot m(\kappa)}
\ee 
or, equivalently,
\be\label{l+m1}
\ell(\tau) + m(\kappa) = 2\left[\dot m U + \dot\ell V\right] \, . 
\ee 
 
A surprising feature of this `dual' description of the Lagrangian density $\LL(U,V)$ of a self-dual NLED is that the new one-variable function $m$ is same as the one-variable {\sl Hamiltonian} function $\frak{h}(\sigma)$!  
This can be seen as follows: replacing $m(\kappa)$ by $\frak{h}(\sigma)$ in \eqref{ell-m} and \eqref{kaptau} we get 
precisely the relations that determine $\frak{h}(\sigma)$ in terms of $\ell(\tau)$, and {\it vice versa}. Furthermore, we know how the functions $\ell(\tau)$ and $\frak{h}(\sigma)$ are related (by a Legendre transform in terms of the variables $\sqrt{2\tau}$ and $\sqrt{2\sigma}$), so $\ell$ and $m$ are related in the same way, which is 
\be\label{l+m2}
\ell(\tau) + m(\kappa) = 2\kappa \dot m = 2\tau \dot\ell\, . 
\ee
Comparing this with \eqref{l+m1} we recover the equations for $\kappa$ and for $\tau$ in terms of $(U,V)$. 

Returning to \eqref{gensol2}, let us replace $m(\kappa)$ by $\frak{h}(\sigma)$, since they are the same function, and 
then relabel the independent variables of the function $\LL$ as follows:
\be\label{cov}
(U,V) \to (v,-u)\,  . 
\ee
We then get 
\be
-\LL(v,-u) = \frak{h}(\sigma) + \frac{2u}{\frak{h}^\prime} \, , \qquad \sigma = v - \frac{u}{(\frak{h}^\prime)^2}\, .
\ee
Comparison with \eqref{generalh} shows that $\HH(u,v)$ is the same function 
as  $-\LL(v,-u)$. Explicitly, given any Lagrangian density $\LL(U,V)$, we may find its Legendre dual
Hamiltonian density $\HH(u,v)$ by the following procedure: 
\be
\label{LtoH}
\LL(U,V)\ \ \longrightarrow \ \ -\LL(v,-u)= \HH(u,v)\ ,
\ee 
Given any Hamiltonian density $\HH(u,v)$ we can similarly find its Legendre-dual Lagrangian density $\LL(U,V)$: 
\be
\label{HtoL}
\HH(u,v)\ \ \longrightarrow \ \ -\HH(-V,U)= \LL(U,V)\ .
\ee

Notice that the change of variables \eqref{cov} implies that 
\be 
V-U \to -(v+u) \, , \qquad V+U \to v-u\, .
\ee
Using the expressions given in the Introduction for $(U,V)$ in terms of $(S,P)$, and $(u,v)$ in terms of $(s,p)$, we deduce that 
\be
S \to -s \, , \qquad \Phi  \to \varphi \, . 
\ee
For Maxwell, for example, we get $\HH = -\LL(s) =s$, as expected. More generally, once $\LL$ is expressed in the form $\LL(S,\Phi)$ we get $\HH$ in the form $\HH(s,\varphi)$ by the following procedure:
\be\label{HLeg}
\boxed{\LL(S,\Phi)\ \ \longrightarrow \ \ -\LL(-s,\varphi)= \HH(s,\varphi)}\ ,
\ee
which is the boxed equation \eqref{magic} of the
Introduction. The converse formula is
\be
\boxed{\HH(s,\varphi)\ \ \longrightarrow \ \ -\HH(-S,\Phi)= \LL(S,\Phi)}\ .
\ee
As we shall see in the following section, these results enormously simplify the task of finding the Hamiltonian density associated to any known Lagrangian density of a self-dual NLED, and {\it vice versa}. 


\subsection{Further examples}

\subsubsection*{No maximum-$\tau$ case}

This possibility was illustrated in \cite{Russo:2024llm} by the choice  $\ell = T (1+ 2e^\gamma\tau/(3T) )^{\frac32}$ which is defined for all $\tau\ge0$ and satisfies the causality conditions of \eqref{causal-L}. The
corresponding Lagrangian density is 
\begin{equation}
\mathcal{L}=\sqrt{2}\, T \left(1+\frac{2e^\gamma V}{3T}-\frac{\Delta}{2} \right) \sqrt{1+\frac{2e^\gamma V}{3T}+\Delta}\, ,  
\end{equation}
with $\Delta= \sqrt{\left(1+\frac{2e^\gamma V}{3T}\right)^2 + 8e^{-\gamma} \frac{U}{3T}}$. 
As expected, it is defined in the entire positive $(U,V)$ quadrant, and it reduces to ModMax
with coupling constant $\gamma$ in the weak-field limit. 

The function $\hh$ for this case is 
\be
\hh=\frac{2}{\sqrt{3} }\, \sqrt{\frac{e^{-\gamma} T}{\sigma}(\Lambda-1)}\left( \sigma-\frac38 e^{\gamma} T(1+\Lambda)\right)
\ ,\qquad 
\Lambda=\sqrt{1+8e^{-\gamma}\sigma/(3T)}\, . 
\ee
The weak-field expansion is $\hh={\rm const.}+e^{-\gamma}\sigma+ O(\sigma^2)$, as expected. 

The standard method of computing the Hamiltonian as a Legendre transform of $\LL $ leads to  complicated equations.
The Courant-Hilbert construction of the Hamiltonian based on \eqref{generalh}
using the above $\hh(\sigma )$ also leads
to complicated equations.
However, the Hamiltonian can be immediately written down by using \eqref{LtoH}. This gives
\be
\HH=\sqrt{2}\, T \left(-1+\frac{2e^\gamma u}{3T}+\frac{\Delta'}{2} \right) \sqrt{1-\frac{2e^\gamma u}{3T}+\Delta'}\, ,
\ee
with
$\Delta' =\sqrt{\left(1-\frac{2e^\gamma u}{3T}\right)^2 + 8e^{-\gamma} \frac{v}{3T}}$.

\subsubsection*{Logarithmic self-dual electrodynamics} 

The choice $\ell =-e^\gamma T \log(1- \tau/T)$ yields the Lagrangian 
density \cite{Russo:2024llm}
\be
\LL = -e^{\gamma } (\Sigma_0-T) -e^{\gamma } T \log
   \left(\frac{e^{2\gamma}}{2U}(\Sigma_0- T)\right)\ ,
   \label{logcero}
\ee
where
\be
\Sigma_0=\sqrt{T^2+4 e^{-2 \gamma } U(T-V) }\ .
\ee
The corresponding $\hh$-function is
\be
\label{hhmm}
\hh= T e^\gamma  \left( M-1-\log \frac{ 1+M}{2}\right) \, , \qquad 
M=\sqrt{1+\frac{4 e^{-2 \gamma }  \sigma }{T}}\, . 
\ee
As in the previous case, in order to obtain the Hamiltonian, one can circumvent the long calculation through a Legendre transform and directly make use of \eqref{LtoH}.
This gives
\be
\HH = e^{\gamma } (\Sigma_0'-T) +e^{\gamma } T \log
   \left(\frac{e^{2\gamma}}{2v}(\Sigma_0'- T)\right)\ ,
   \label{logcero}
\ee
where
\be
\Sigma_0'=\sqrt{T^2+4 e^{-2 \gamma } v(T+u) }\ .
\ee

\section{$\Phi$-parity duality}\label{sec:ND}

The self-duality PDE \eqref{PDEL} is invariant under the ``$\Phi$-parity''  transformation $\Phi\to -\Phi$, which therefore takes a given solution $\LL(S,\Phi)$ into its $\Phi$-parity dual solution 
\be
\hat \LL(S,\Phi) =\LL(S,-\Phi)\ .
\ee
The Hamiltonian density $\hat\HH(s,\varphi)$ corresponding to $\hat\LL(S,\Phi)$ can therefore be found by using the formula \eqref{HLeg}:
\be\label{hrela}
\LL(S, -\Phi)\ \longrightarrow \ - \LL(-s, -\varphi) = \HH(s,-\varphi)\, , 
\ee
and hence
\be
\hat\HH(s,\varphi) = \HH(s, -\varphi) \, . 
\ee
Obviously, $\hat\LL=\LL$ whenever $\LL(S,\Phi)$ is $\Phi$-parity invariant; in this class of NLED theories the weak-field expansion of $\LL(S,\Phi)$ is a power-series expansion in $S$ and $P^2$. Otherwise (i.e. $\hat\LL \neq \LL$) both have a weak-field expansion in powers of $S$ and $\Phi$, with odd powers of $\Phi$ that cannot be rewritten as a sum of positive powers of $S$ and $P^2$. 

The transformation $\Phi\to -\Phi$ for fixed $S$ is equivalent to 
\be
\label{UVduality}
(U,V) \to -(V,U)\, . 
\ee
Similarly, the transformation $\varphi\to -\varphi$ for fixed $s$ is equivalent to 
\be
(u,v) \to (v,u)\, . 
\label{uvduality}
\ee
The formula \eqref{hrela} connects $\hat \HH(u,v)$ to the original Lagrangian
density $\LL$.  Using the version of this formula for $\LL(U,V)$, i.e. \eqref{LtoH}, we have 
\be\label{LL-HH}
\hat\LL(U,V) = \LL(-V,-U) \longrightarrow - \LL(u,-v) = \hat\HH(u,v)\, . 
\ee
Similarly, applying \eqref{HtoL} to the $\varphi$-parity dual of $\HH(u,v)$ yields 
\be\label{HH-LL}
\hat\HH(u,v) = \HH(v,u)\ \longrightarrow \ - \HH(U,-V) = \hat\LL(U,V)\, . 
\ee

These results show how the Lagrangian and Hamiltonian densities of generic NLED theories are related to those of their $\Phi$-parity duals.  We now turn to the special class of (electromagnetic) self-dual NLED theories.

\subsection{The $\Phi$-parity dual of self-dual NLED}

Using the CH constructions for $\LL$ and $\hat\HH$ in \eqref{LL-HH} we see that 
\be
- \ell(\tau) +\frac{2u}{\dot\ell(\tau)} = \hat\hh(\sigma) +  
\frac{2u}{[\hat\hh^\prime](\sigma) }\, ,
\ee
with
\be
\tau= -v + \frac{u}{\dot\ell^2(\tau)}\, , \qquad \sigma= v - \frac{u}{[\hat\hh^\prime]^2(\sigma)} \, .   
\ee
These equations imply that 
\be
\hat\hh(\sigma) = - \ell(\tau) \, , \qquad \tau=-\sigma \qquad
\left[\Rightarrow\ \hat\hh^\prime(\sigma) = \dot\ell(\tau)\right]\, , 
\ee
and hence that 
\be \label{box1}
\boxed{\ell(-\sigma) = -\hat\hh(\sigma)}\, . 
\ee
Similarly, using the CH constructions for $\HH$ and $\hat\LL$ in \eqref{HH-LL}, we conclude that
\be\label{box2}
\boxed{\hh(-\tau) =- \hat\ell(\tau)} \, . 
\ee

\begin{figure}[h!]
 \centering
\includegraphics[width=0.7\textwidth]{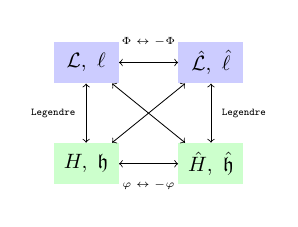}
  \caption{Schematic picture of different transformations.
  The cross arrows represent the simple relations
  $\hat\hh(\sigma )=-\ell(-\sigma)$ and $\hat\ell(\tau )=-\hh(-\tau)$. For theories symmetric under $\Phi\to -\Phi$, $\hat\LL =\LL$ and  $\hh(\sigma)=-\ell(-\sigma)$.
 }
 \label{fig5}
\end{figure}

We mentioned in the Introduction that the existence of a weak-field limit implies that $\ell(\tau)$ is defined for $\tau<0$, even though only its values for $\tau\ge 0$ are relevant to the CH formula for $\LL(U,V)$. Now we see from \eqref{box1} that the function $\ell(\tau)$ for $\tau\le0$ is (minus) the one-parameter function $\hat\hh$ for the ``$\varphi$-parity'' dual of the Hamiltonian density $\HH$ that is Legendre-dual to $\LL$. Similarly, we see from  \eqref{box2} that the function $\hh(\sigma)$ for $\sigma\le0$ is (minus) the one-parameter function $\hat\ell$ for the ``$\Phi$-parity'' dual of the Lagrangian density $\LL$. The general picture is illustrated in fig. \ref{fig5},
and we present some illustrative examples below.

For the special case of $\Phi$-parity invariant theories, $\hat\ell=\ell$ and $\hat\hh =\hh$ and the two relations \eqref{box1} and \eqref{box2}
reduce to the one relation 
\be\label{kinR}
\boxed{\ell(\kappa)+ \hh(-\kappa)=0\, ,  \quad \kappa \in \bb{R}}\, . 
\ee
Born-Infeld provides a simple example, and we return to a study of $\Phi$-parity invariant self-dual NLED theories at the end of this section.  

We now present examples that illustrate the general case. 

\subsection*{Illustrative examples}

Let us consider self-dual NLED theories defined by 
\be
\ell(\tau ) = -T\left(1-\frac{e^{\gamma }\tau}{qT}\right)^{q }\ .
\ee
According to the formula \eqref{box1}, the $\hh$-function of the $\Phi$-dual theory is 
\be
\hat \hh(\sigma)=-\ell(-\sigma) = T\left(1+\frac{e^{\gamma }\sigma}{qT}\right)^{q }\ .
\ee
For $\gamma=0$ this is the same as the ``$q$-deformed" function of \eqref{hhqqq},  which tells us that the case under consideration now is, for $\gamma=0$, the $\Phi$-parity dual of the ``$q$-deformed"
case of subsection \eqref{sec:examples}

For $q=1/2$ we have the ModMaxBorn theory. In this case 
\be
\begin{aligned} 
\ell_{\rm MMB}(\tau) &=\  - \sqrt{T^2-2e^\gamma T\tau} \\
\hh_{\rm MMB}(\sigma) &= \sqrt{T^2 +2e^{-\gamma}T\sigma}\, , 
\end{aligned}
\ee
but $\Phi$-duality flips the sign of $\gamma$, so that 
\be
\begin{aligned}
\hat\ell_{\rm MMB}(\tau) &=\  - \sqrt{T^2-2e^{-\gamma} T\tau} \\ 
\hat\hh_{\rm MMB}(\sigma) &= \sqrt{T^2 +2e^{\gamma}T\sigma}\, , 
\end{aligned}
\ee
and both \eqref{box1} and \eqref{box2} are therefore satisfied. 
Obviously, in the BI ($\gamma=0$) case there is no distinction
between the dual (hatted) functions and the original functions, since BI theory is $\Phi$-parity invariant.

\subsubsection*{The $q=3/4$ case}

The Lagrangian and Hamiltonian densities can also be found explicitly in the $q=3/4$ case, where
\be\label{lll}
\ell(\tau) = -T\left(1-\frac{4e^{\gamma }\tau}{3T}\right)^{\frac34 }\ .
\ee
Using \eqref{gensol}, we find 
\be
\LL = - T \left(\Lambda - \frac{2e^{-\gamma}U}{3T}\right)^{\frac12} 
\left( \Lambda + \frac{4e^{-\gamma}U}{3T}\right)\, , 
\label{letres}
\ee
where 
\be
\Lambda= \sqrt{1- \frac{4Ve^\gamma }{3T} + \left(\frac{2e^{-\gamma}U}{3T}\right)^2}\, . 
\ee

Using the relations \eqref{dlph}, \eqref{sigtau}, \eqref{ell+h} we find the 
corresponding $\hh$-function:
\be
\label{hhdf}
\hh(\sigma) =T \left(\sqrt{1+\frac{4 e^{-2 \gamma }\sigma^2}{9 T^2}}+\frac{4e^{-\gamma }\sigma}{3T}\right) \sqrt{\sqrt{1+\frac{4 e^{-2 \gamma }\sigma^2}{9 T^2}}-\frac{2e^{-\gamma }\sigma}{3T}}\, .  
\ee
The  Hamiltonian can now be found via the CH construction of \eqref{generalh}, but it is  much simpler to use \eqref{LtoH} to obtain
\be
\HH =  T \left(\Lambda' - \frac{2e^{-\gamma}v}{3T}\right)^{\frac12} 
\left( \Lambda' + \frac{4e^{-\gamma}v}{3T}\right)\, , 
\label{letres}
\ee
where 
\be
\Lambda'= \sqrt{1+ \frac{4ue^\gamma }{3T} + \left(\frac{2e^{-\gamma}v}{3T}\right)^2}\, . 
\ee

Notice that $\hat\hh(\sigma)$ of \eqref{hhdf} is different from $-\ell(-\sigma)$ of \eqref{lll}, even at $\gamma =0$. This tells us that this theory is {\sl not} $\Phi$-parity self-dual, even at $\gamma =0$. The one-parameter CH functions of the $\Phi$-parity dual theory are easily found  from \eqref{box1} and \eqref{box2}:
 \be\label{hra}
\hat \hh =T \left(1+\frac{4e^{\gamma }\sigma}{3T}\right)^{\frac34}\, , 
 \ee
\be
\label{lldf}
\hat \ell(\tau) =-T \left(\sqrt{1+\frac{4 e^{-2 \gamma }\tau^2}{9 T^2}}-\frac{4e^{-\gamma }\tau}{3T}\right) \sqrt{\sqrt{1+\frac{4 e^{-2 \gamma }\tau^2}{9 T^2}}+\frac{2e^{-\gamma }\tau}{3T}}
  \, .  
\ee
Comparing with the example \eqref{hhqqq} for $q=3/4$, we note that $\Phi$-duality has again flipped the sign of $\gamma$; for weak fields the $\Phi$-parity dual theory becomes ModMax but with $\gamma\to -\gamma $.

The $\Phi$-dual Lagrangian $\hat \LL$ and Hamiltonian $\hat \HH $ can now be found by the maps \eqref{UVduality}, \eqref{uvduality}. Alternatively, they can be obtained
from the CH construction using the above expressions for $\hat\ell$ and $\hat\hh$. For example, from \eqref{generalh} and \eqref{hra}, we have the equations
\be
\sigma+e^{2\gamma}u\sqrt{1+\frac{4e^{\gamma}\sigma  }{3T}}-v=0\ .
\ee
This gives
\be
\sqrt{1+\frac{4e^{\gamma}\sigma  }{3T}}=-\frac{2e^{-\gamma}u}{3T}+\Sigma\ ,\qquad \Sigma =\sqrt{1+\frac{4ve^{\gamma} }{3T} +\left(\frac{2e^{-\gamma}u}{3T}\right)^2}\, , 
\ee
which leads to 
\be
\label{aann}
\hat \HH = T\left(\Sigma +
\frac{4e^{-\gamma}u}{3T}\right)\left(\Sigma -
\frac{2e^{-\gamma}u}{3T}\right)^{\frac12}\, , 
\ee
in accordance with the formula $\hat \HH(u,v)=\HH(v,u)$.

\subsection{The alternative CH construction}

In addition to the  construction of \eqref{gensol} that gives the Lagrangian density $\LL$ of the general self-dual NLED in terms of a boundary function $\ell$, Courant and Hilbert show that the solution to the partial differential equation \eqref{GR} may also be expressed as \cite{C&H}
\be
\label{ddos}
\LL(U,V) =\frac{V}{x}- x U +\omega(x)\ , 
\ee
where $\omega(x)$ is defined for positive {\sl dimensionless} variable $x$, which is determined implicitly by the equation
\be
\label{duno}
x\omega'(x) = xU + \frac{V}{x}\ .
\ee
To verify this we take the differentials of both sides of \eqref{ddos}. Simplifying the result by using \eqref{duno} we find that 
\be
d\LL = \frac{dV}{x} - x dU \, , 
\ee
and hence that $\LL_U\LL_V =-1$. We also see, by comparison with \eqref{partials1},  that the relation of $\omega(x)$ to $\ell(\tau)$ must be such that $\dot\ell =1/x$. In fact, the relation is given implicitly by 
\be\label{ellomega}
\ell(\tau) = \omega(x) + x\omega^\prime(x)\, , \qquad\tau= x^2 \omega^\prime(x)\, , 
\ee
from which we find that 
\be
\dot \ell = \left(2\omega^\prime + x^2 \omega^\prime{}^\prime\right)\frac{dx}{d\tau}  
= \frac{1}{x} \frac{d (x^2\omega^\prime)}{dx} \frac{dx}{d\tau} = \frac{1}{x}\, , 
\ee
as expected. This alternative to the CH constructions described in the Introduction is useful when considering the implications of $\Phi$-parity; conversely,
consideration of $\Phi$-parity yields insights into the relation between $\ell$ and $\omega$ that are in some respects similar to what we have already found for $\ell$ and $\hh$.  

Recall that $\Phi\to-\Phi$ is equivalent to $(U,V)\to -(V,U)$. Applying this to \eqref{ddos} we find that the $\Phi$-parity transform of $\LL(U,V)$ is 
\be\label{ddos2}
\hat \LL(U,V) = \ x V - \frac{U}{x} + \omega(x)
\ee
where $x$ is now determined by the equation
\be\label{duno2}
-x\omega'(x) = xV + \frac{U}{x}\ .
\ee
If we now define a new variable $y$ and a new function 
$\hat\omega$ by 
\be 
y := \frac{1}{x} \, , \qquad 
\hat\omega(y) := \omega(x)\, , 
\ee
then the equations \eqref{ddos2} and \eqref{duno2} 
defining $\hat\LL(U,V)$ become, respectively
\be
\label{ddoshat}
\hat\LL(U,V) =\frac{V}{y}- y U +\hat\omega(y)\ , 
\ee
and 
\be\label{duno22}
y\hat\omega'(y) = yU + \frac{V}{y}\ .
\ee
These are {\sl formally} the same as the original equations
\eqref{ddos} and \eqref{duno} that define $\LL(U,V)$,
but the function $\hat\omega$ determining $\hat\LL$ is 
generally different from the function $\omega$ determining $\LL$, since $\hat\omega(x) = \omega(1/x)$. In the following 
subsection we focus on the special class of $\Phi$-parity
invariant theories for which $\omega(x)=\omega(1/x)$, and hence $\hat\LL=\LL$.

There is also a CH construction of $\hat\LL$ in terms of a 
function $\hat\ell$, with a relation of $\hat\ell$
to $\hat\omega$ that is formally the same as the relation 
of $\ell$ to $\omega$ expressed by the equations of \eqref{ellomega}. The relation of $\ell$ to 
$\hat\omega$ is different, however. In terms of the new variable $y$ and the new function $\hat\omega$, the equations of \eqref{ellomega} become
\be\label{ellOmega}
\ell(\tau) = \hat\omega(y) + \tau y \, , \qquad 
\tau = - \hat\omega^\prime(y)\, .
\ee
This has a remarkably simple interpretation: it 
tells us that $\ell(\tau)$ is the Legendre transform of 
$-\hat\omega(y)$ with respect to $y$:
\be\label{LTagain1}
\ell(\tau) = \sup_y \left\{y\tau + \hat\omega(y)\right\}\, .
\ee

From the equation for $\tau$ in \eqref{ellOmega} we have 
\be\label{convex}
-\hat\omega^\prime{}^\prime(y) =  \frac{d\tau}{dy} = - x^2 \frac{d\tau}{dx} \, , 
\ee
which is the inverse of $\ddot\ell$ (since $\dot\ell=1/x$); i.e.
\be\label{dds}
\ddot\ell(\tau) \hat\omega^\prime{}^\prime(y) =-1\, . 
\ee
This result tells us that $\hat\omega(y)$ is a strictly concave function iff $\ell(\tau)$ is a strictly 
convex function, as is required for causality, except that causality also allows $\ddot\ell=0$, 
which is realized by ModMax and its Maxwell limit. These conformal NLED theories are therefore 
not obviously included in the alternative CH construction of self-dual NLED theories. 

To better understand why ModMax and Maxwell are special cases, we observe that the converse of \eqref{LTagain1} is 
\be\label{LTagain2}
- \hat\omega(y) = \sup_\tau \left\{ y\tau - \ell(\tau)\right\}\, .
\ee
That is, $-\hat\omega(y)$ is the Legendre transform of 
$\ell(\tau)$, {\sl with respect to $\tau$}
(recall that $\hh(\sigma)$, expressed as the function $H(\sqrt{2\sigma})$, is its Legendre transform with respect to $\sqrt{2\tau}$). 
For the choice $\ell(\tau)= e^\gamma \tau$, we have
\be\label{defhato}
\hat\omega(y) = \sup_\tau \left\{\left (y- e^\gamma\right)\tau\right\}\, , 
\ee
which is defined only for $y= e^\gamma$, and is zero at this one point in its domain. Using this function in \eqref{ddos2} yields the ModMax Lagrangian density, and Maxwell for $\gamma=0$. Thus, \eqref{ddoshat} does include Modmax and Maxwell if the function $\hat\omega$ is defined in terms of the CH function $\ell$, as in \eqref{defhato}, and a similar (dual) statement 
applies if the function $\omega$ in \eqref{ddos} is defined as (minus) the Legendre transform of $\hat\ell$.

One utility of the alternative CH construction described above is that new explicit examples of self-dual NLED theories can be found that would otherwise be difficult to find. This is illustrated by the following example.

\subsubsection*{Generalized Logarithmic NLED}

We start from the function 
\be\label{genlog}
\omega(x)= c T -\frac{T}{2}\left( e^\gamma x + \frac{1}{e^\gamma x}\right) +\eta T\log(x)\ , 
\ee
where $c$ is a parameter that can be  chosen to arrange for zero vacuum energy, 
and $\eta$ is a further real parameter. Using \eqref{ddos} we find the Lagrangian density:
\be
\LL =cT-\Sigma-\eta T \log\left(\frac{\Sigma-\eta T}{e^\gamma T+U} \right)\ ,
\ee
with
\be
\Sigma\equiv \sqrt{(T+2e^{-\gamma}U)(T-2e^{\gamma}V)+\eta^2 T^2}\ .
\ee
For $\eta=0$ this reduces to ModMaxBorn.

Recalling again that $\Phi$-parity takes $(U,V)$ to $-(V,U)$, one sees from inspection 
that this generalized logarithmic NLED theory is $\Phi$-parity invariant iff  $\eta=0$ 
and $\gamma=0$, in which case it reduces to Born-Infeld.  For all other choices of these 
parameters the $\Phi$-parity dual is found by changing the signs of both $\eta$ and $\gamma$.

\subsection{The general ``$\Phi$-parity" invariant self-dual theory}

Comparing \eqref{ddos2} to \eqref{ddoshat} we see that $\hat\LL = \LL$ whenever $\hat\omega= \omega$, i.e. whenever $\omega(1/x) = \omega(x)$. 
In this case
\be
\omega^\prime(x) = - \frac{1}{x^2} \omega^\prime (1/x) \, , 
\ee
which implies that $\omega^\prime(1)=0$. From the equation for $\tau$ in \eqref{ellomega} we
see that $x=1$ is equivalent to $\tau=0$, so the weak-field expansion 
\be\label{BIell}
\ell(\tau) = \tau + \frac{1}{2T}\tau^2 + \mathcal{O}(\tau^3)
\ee
must be equivalent to an expansion of $\omega(x)$ about $x=1$. From the expressions for 
$\ell(\tau)$ and $\tau$ in \eqref{ellomega}, one finds that this expansion is 
\be
\omega(x) = - \frac{T}{2} (1-x)^2 + \mathcal{O}[(1-x)^3]\, . 
\ee 
This result is a direct consequence of the fact that $\omega^\prime(1)=0$ and the identity 
\eqref{dds} (for $\tau=(1-x)=0$).  The corresponding weak-field expansion of $\LL$ is 
\be
\LL(S,\Phi) = S + \frac{\Phi^2}{2T} + \mathcal{O}(1/T^2)\, . 
\ee

A very simple choice for $\omega(x)$ that is manifestly invariant under $x\to 1/x$ is
\be\label{BIx}
\omega(x) =  T -\frac{T}2\   \big(x+ x^{-1}\big) \, .
\ee
In this case the solution of \eqref{duno} for $x$ is
\be
x= \sqrt{\frac{T-2U}{T+2V}}  \, . 
\ee
This yields the Born-Infeld theory. The weak-field expansions of  $\omega$ and $\ell$ are exactly as above
in this case. In general, a rescaling of the parameter $T$ in \eqref{BIell} and \eqref{BIx} will be necessary. 

It is a simple matter to write down other functions $\omega(x)$ that are invariant under $x\to1/x$, but any such function 
must also have the property that the equation \eqref{duno}  has a unique solution for $x$, 
and we must also impose causality conditions. For example, the condition $\dot\ell\ge1$ requires $x\le 1$.  
Another aid to separating the causal from the acausal NLED theories is the relation \eqref{dds}, 
which implies that $\hat\omega(y)$ is a concave function of $y$ whenever $\ell(\tau)$ is a convex function of $\tau$, as required. 
The implications for $\omega(x)$ are generically not obvious but  $\hat\omega= \omega$ for $\Phi$-parity invariant theories, and therefore 
$\omega(x)$ must also be concave (for $x\le1$). 

Consider, for example the following one-parameter generalization of Born-Infeld, defined by
\be\label{BI+}
\omega (x) =- \frac{T}{2}\left\{ \left(x+ \frac{1}{x}\right) + a \left(x+ \frac{1}{x}\right)^2\right\}\, , 
\ee
where $a$ is a constant. One finds that 
\be\label{prime}
\omega^\prime(x) = T\left\{ \frac{(1-x^2)}{2x^2} + a\frac{(1-x^4)}{x^3}\right\}
\ee
and 
\be\label{primeprime}
\omega^\prime{}^\prime(x) = 
- \frac{T}{x^4}\left(x+ ax^4 +3a\right)\, .  
\ee
Using \eqref{prime} in \eqref{duno} we find the following equation for $x$:
\be\label{xeq}
(T-2V) + \frac{2aT(1-x^4)}{x} =  (T+2U) x^2\, . 
\ee
Inspection of the graphs of the functions of $x$ on both sides of this equation shows that a unique solution exists for all $(U,V)$ in the positive quadrant iff $a\ge0$. 
From \eqref{primeprime} we see that this is also required for $\omega(x)$ to be a concave function for $0\le x\le1$.
For $a>0$ we have a  one parameter self-dual deformation of Born-Infeld that preserves $\Phi$-parity invariance. Moreover, 
as \eqref{xeq} is a quartic equation for $x$, which has an explicit and unique solution for $0\le x\le 1$, the Lagrangian density can still be found explicitly.

\section{Legendre self-duality}

So far we have considered the Lagrangian and Hamiltonian densities. In both cases, the NLED field equations
may be written in first-order form as the ``macroscopic Maxwell equations''
\be
\begin{aligned}
\dot{\bf D} =  \boldsymbol{\nabla}\times {\bf H}  \ \ &\, , \qquad \boldsymbol{\nabla}\cdot {\bf D} = 0\, , \\
\dot {\bf B} =   - \boldsymbol{\nabla}\times {\bf E} &\, , \qquad \boldsymbol{\nabla}\cdot {\bf B}=0\, , 
\end{aligned}
\ee
together with {\sl constitutive relations} which are  either
\be
{\bf D}  = \partial{\LL}/\partial {\bf E} \, , \qquad {\bf H} = -\partial{\LL}/\partial {\bf B}\, ,  
\ee
or 
\be
{\bf E}  = \partial{\HH}/\partial {\bf D} \, , \qquad {\bf H} = \partial{\HH}/\partial {\bf B}\, . 
\ee
However, we may also specify the constitutive relations in terms of  a `dual' Hamiltonian density 
\be
\tilde \HH({\bf E}, {\bf H}) = \sup_{\bf B}\left\{ -{\bf B} \cdot{\bf H} - \LL\right\} \, ,
\ee
in which case 
\be
{\bf D} = -\frac{\partial \tilde\HH}{\partial {\bf E}} \, , \qquad {\bf B} = -\frac{ \partial\tilde \HH}{\partial {\bf H}}\, , 
\ee
or in terms of a `dual' Lagrangian density 
\be\label{dualL}
\tilde \LL({\bf D}, {\bf H}) = \sup_{({\bf E},{\bf B})} \left\{ \LL - {\bf E}\cdot {\bf D} + {\bf B}\cdot {\bf H}\right\}\, , 
\ee
in which case 
\be
{\bf E} = -\frac{\partial \tilde\LL}{\partial {\bf D}}\, , \qquad {\bf B} =  \frac{\partial \tilde \LL}{\partial {\bf H}}\, . 
\ee
The possibility of a description in terms of one of four ``fundamental functions'' was observed by Born in \cite{Born:1937drv} but here we use the 
more standard terminology\ of Bialynicki-Birula \cite{Bialynicki-Birula:1984daz}, except for sign changes to ensure that the addition of a constant to $\LL$ 
implies the addition of the same constant to $\tilde \LL$, and 
its subtraction from both $\HH$ and $\tilde\HH$. 

Another of Born's observations was that, for Born-Infeld, $\LL$ and $\tilde \LL$ are  identical functions of their respective scalar variables, appropriately defined; this has been called ``Legendre self-duality''. As mentioned in the Introduction, this was shown by Gaillard and Zumino to be a property of any 
(electromagnetically) self-dual NLED theory \cite{Gaillard:1997rt}, and a later proof of Theisen and Kuzenko \cite{Kuzenko:2000zw} showed that only a $Z_2$ electromagnetic duality was needed. The starting point
of this proof was (in our notation) the Lagrangian density
\be 
\LL(F,\tilde A) = \LL(F) - {\bf B}\cdot \tilde{\bf E} - {\bf E}\cdot  \tilde{\bf B}\, , 
\ee
where $({\bf E},{\bf B})$ are the electric/magnetic components of $F$, now an arbitrary 2-form field, and 
$(\tilde{\bf E},\tilde{\bf B})$ are the electric/magnetic components of $\tilde F = d\tilde A$. The combined
field equations found from varying both $F$ and $\tilde A$  are equivalent to those of $\LL(F)$ for  $F=dA$ 
(since variation of $\tilde A$ yields the equation $dF=0$). However, the equations found from varying $F$, which are
\be\label{TK}
\tilde {\bf B} =\frac{\partial \LL(F)}{\partial {\bf E}} = {\bf D}\, , \qquad 
\tilde {\bf E} = \frac{\partial \LL(F)}{\partial {\bf B}} = - {\bf H}\, , 
\ee
may be used to eliminate $F$;  this yields the dual Lagrangian density $\tilde \LL(\tilde F)$.  Theisen and Kuzenko show that the functions $\LL(F)$  and $\tilde \LL(\tilde F)$ are the same for all NLED invariant under a discrete $Z_2$ electromagnetic duality transformation. To state this result in our notation, we observe that 
a further implication of the equations \eqref{TK} is 
\be\label{app}
\begin{aligned}
\tilde S =&\  \frac12\left(|\tilde {\bf E}|^2 - |\tilde {\bf B}|^2\right) \ \equiv  -\frac12\left(|{\bf D}|^2- |{\bf H}|^2\right) \, , \\
\tilde P =&\ \tilde{\bf E}\cdot\tilde{\bf B} \ \equiv - {\bf D}\cdot{\bf H}\, .  
\end{aligned}
\ee
In other words, Legendre self-duality can be restated as 
the equivalence, for self-dual NLED,  of the functions $\LL(S,P)$ and $\tilde\LL(\tilde S,\tilde P)$, 
with $(\tilde S,\tilde P)$ defined in terms of $({\bf H},{\bf D})$ according to \eqref{app}. 

Here we show that this result  follows directly from the definition of the dual Lagrangian density $\tilde\LL$ whenever
the Hamiltonian density is invariant under  the $-
\pi/2$ duality-rotation taking  $({\bf D},{\bf B})$ to $({\bf B}, -{\bf D})$,
which implies that
\be\label{sdprop}
\HH({\bf D},{\bf B}) = \HH({\bf B},-{\bf D}) \, . 
\ee
This is obviously a property of any self-dual NLED, but it also a property of some other NLED theories
that are {\sl not} self-dual. 

We begin with the observation that 
\be\label{start}
\HH({\bf D},{\bf B}) = \sup_{\bf E}\left\{ {\bf D} \cdot {\bf E} - \LL({\bf E}, {\bf B}) \right\} \, . 
\ee
This relation implies the following two relations 
\be\label{LLs}
\begin{aligned}
\LL({\bf E}, {\bf B}) =&\  \sup_{\bf D} \left\{ {\bf E} \cdot {\bf D} - \HH({\bf D},{\bf B}) \right\} \, , \\
\tilde\LL ({\bf D},{\bf H}) =& \ \sup_{\bf B} \left\{ {\bf H}\cdot {\bf B} - \HH({\bf D},{\bf B}) \right\} \, .
\end{aligned}
\ee
The first of these is just the inverse of \eqref{start}. The second follows by using \eqref{start} 
to replace $\HH$ on the right-hand side; this yields the definition  of \eqref{dualL} for $\tilde\LL$.
We see from these relations that both $\LL$ and $\tilde\LL$ are Legendre transforms of $\HH({\bf D},{\bf B})$, 
but one is with respect to the first 3-vector variable and the other is with respect to the second 3-vector variable. 
Let us now rewrite \eqref{LLs} more abstractly as 
\be
\begin{aligned}
\LL({\bf X},{\bf Y}) =& \  \sup_{\bf Z} \left\{ {\bf X} \cdot {\bf Z} - \HH({\bf Z},{\bf Y})\right\}\, ,  \\
\tilde\LL({\bf Y},-{\bf X}) =& \  \sup_{\bf Z} \left\{  {\bf X} \cdot {\bf Z} - \HH({\bf Y},-{\bf Z})\right\} \, , 
\end{aligned}
\ee 
for 3-vectors $({\bf X},{\bf Y},{\bf Z})$.  From this, we see that the property \eqref{sdprop} 
of any self-dual theory implies that 
\be
\LL({\bf X},{\bf Y}) = \tilde\LL({\bf Y},-{\bf X}) \, . 
\ee
Given Lorentz invariance of the function  $\LL({\bf X},{\bf Y})$ on the left-hand side, it may be expressed as a function of the Lorentz scalars
$|{\bf X}|^2- |{\bf Y}|^2$ and  ${\bf X}\cdot {\bf Y}$. The same is true of the right-hand side but with $(X,Y)$ replaced by $(Y,-X)$, as we should expect from the minus signs in the definitions of $(\tilde S,\tilde P)$ in \eqref{app}.  
We thus conclude that $\LL(S,P)$ and $\tilde\LL(\tilde S,\tilde P)$  are the same, as functions, for any self-dual NLED theory. 

As a simple illustration of Legendre self-duality, 
we consider a class of NLED theories, introduced in \cite{Kruglov:2009he}, that may be defined by the following one-parameter family of Lagrangian densities  in which $(a,b)$ are a pair of auxiliary fields: 
\be\label{RTL}
\LL_{\rm RT}  =  T-  \frac{T}{2} \left[ a+ \frac{(1+b^2)}{a}\right]  + aS + \xi bP\, . 
\ee
The family parameter is $\xi$, which we may assume to be non-negative. For $\xi=1$ we have the Ro{\v c}ek-Tseytlin  (RT) formulation of Born-Infeld \cite{Rocek:1997hi}; elimination of the auxiliary fields yields $\LL_{\rm BI}(S,P)$.  For $\xi=0$, we get the original Born theory 
\cite{Born:1937drv}; the general case was discussed in detail  in \cite{Russo:2024kto}.  An advantage of this auxiliary-field formulation is that the 3-vector fields $({\bf D},{\bf H})$ are now linear functions of $({\bf E},{\bf B})$:
\be\label{DH-EB}
\left(\begin{array}{c}  {\bf D} \\ {\bf H} \end{array}\right)  = \left(\begin{array}{cc}  a & \xi b  \\ -\xi b & a \end{array}\right) 
\left(\begin{array}{c}  {\bf E} \\ {\bf B} \end{array}\right) \, .
\ee
This implies that 
\be
{\bf E}\cdot {\bf D} - {\bf B}\cdot {\bf H} = 2(a S+ \xi b P)\, ,   
\ee
and hence that the dual RT Lagrangian density is
\be
\tilde \LL_{\rm RT}  = T - \frac{T}{2} \left[ a+ \frac{(1+b^2)}{a}\right]  - (aS +\xi bP)\, , 
\ee
but with $(S,P)$ expressed as functions of $(\tilde S,\tilde P)$.

Using \eqref{app} and \eqref{DH-EB}, it is straightforward to show that 
\be\label{tS-tP}
\left(\begin{array}{c} S \\ P\end{array} \right) =  \frac{1}{(a^2+\xi^2b^2)^2} 
\left(\begin{array}{cc} \xi^2b^2-a^2 & 2\xi ab \\ -2\xi ab& \xi^2 b^2-a^2\end{array} \right) 
\left(\begin{array}{c} \tilde S \\ \tilde P \end{array}\right) \, , 
\ee
and hence that  
\be
aS + \xi bP = -\left(\tilde a \tilde S + \xi \tilde b \tilde P\right) \, ,
\ee
where 
\be
\tilde a = \frac{a}{a^2+\xi^2b^2} \, , \qquad \tilde b = -\frac{b}{a^2+\xi^2 b^2} \, . 
\ee
This auxiliary-field redefinition is such that 
\be
a+ \frac{(1+b^2)}{a} = \tilde a + \frac{(f_\xi(\tilde a,\tilde b) +\tilde b^2)}{\tilde a}\, , 
\ee
where 
\be
f_\xi(\tilde a,\tilde b) = 1+ (\xi^2-1)\tilde b^2 \left[1 - \frac{1}{\tilde a^2 + \xi^2\tilde b^2}\right] 
\ee
We thus deduce for $\xi=1$, that 
\be
\tilde \LL^{(\xi=1)}_{\rm RT}  = T + \frac{T}{2} \left[ \tilde a+ \frac{(1+\tilde b^2)}{\tilde a}\right]  + \tilde a\tilde S +  \tilde b\tilde P\, . 
\ee
As this is formally the same as the $\xi=1$ case of \eqref{RTL}, elimination of the auxiliary fields $(\tilde a, \tilde b)$ now yields the BI Lagrangian density but with $(S,P)$ replaced by $(\tilde S,\tilde P)$; i.e. 
\be
\tilde \LL_{\rm BI}(\tilde S,\tilde P) = T- \sqrt{T^2 -2T\tilde S - \tilde P^2}\, ,  
\ee
which is formally identical to $\LL_{\rm BI}(S,P)$. 

For all other values of $\xi$, we have $\tilde\LL_{\rm RT} \ne \LL_{\rm RT}$, but $\xi=0$ is a special case 
because then the only $\tilde b$-dependence is via the $\tilde b^2$ term of $f_\xi$, which implies that 
$f_\xi\to 1$ upon elimination of $\tilde b$.  Elimination of $\tilde a$ then yields $\LL_{\rm Born}$, so Born's original 
theory is also Legendre self-dual. The reason for this is that  $\HH_{\rm Born}$ satisfies the condition \eqref{sdprop}. 
The results of \cite{Russo:2024kto} for the Hamiltonian density for arbitrary $\xi$ show that \eqref{sdprop} is satisfied 
{\sl only} for $\xi=0$ and $\xi=1$.

\subsection{A proof from the CH formula}

We shall now present a proof that (electromagnetic) self-duality implies Legendre self-duality, 
by taking the CH formula \eqref{gensol} as our starting point.  We know the first 
derivatives of $\LL(U,V)$ from \eqref{partials1}, and we know how 
$(U,V)$ depend on  $(S,P)$ and hence on ${\bf E}$ and ${\bf B}$. This allows us to compute the derivatives 
of $\LL$ with respect to both ${\bf E}$ and ${\bf B}$.  Recalling the definitions of 
$({\bf D}, {\bf H})$ as derivatives of $\LL$, we find that 
\be\label{DH-EB2}
\left(\begin{array}{c}  {\bf D} \\ {\bf H} \end{array}\right)  = \left(\begin{array}{cc}  a &  b  \\ -b & a \end{array}\right) 
\left(\begin{array}{c}  {\bf E} \\ {\bf B} \end{array}\right) \, , 
\ee
where, now, 
\be
a= \frac{\dot\ell^2 V+U}{\dot\ell(V+U)} \, , \qquad b= \frac{(\dot\ell^2-1)P}{2\dot\ell (V+U)}\, . 
\ee
From this result we may compute $(\tilde S,\tilde P)$. One finds that 
\be
\tilde S = \frac{U}{\dot\ell^2} -\dot\ell^2 V \, \qquad \tilde P = -P\, , 
\ee
which allows us to determine $(\tilde U,\tilde V)$ in terms of $(U,V)$ and $\dot\ell$. The result is 
\be
\tilde U = \dot\ell^2 V\, , \qquad \tilde V= \frac{U}{\dot\ell^2} \, .
\ee
We also find from \eqref{DH-EB2} that 
\be
-{\bf E}\cdot {\bf D} + {\bf B}\cdot{\bf H} =   \frac{2U}{\dot\ell} - 2\dot\ell V\, , 
\ee
and hence, using the CH formula for $\LL$, that  
\be
\begin{aligned}
\tilde\LL  &=\ \ell(\tau) - \frac{2U}{\dot\ell} + \left[ \frac{2U}{\dot\ell} - 2\dot\ell V\right] \equiv  \ell(\tau) - 2\dot\ell V \\
&=\  \ell(\tau) - \frac{2\tilde U}{\dot\ell}\, , 
\end{aligned}
\ee
where 
\be
\tau= V + \frac{U}{\dot\ell^2} = \tilde V + \frac{\tilde U}{\dot\ell^2}\, . 
\ee
We thus deduce that $\tilde\LL(\tilde U,\tilde V)$ is given by the CH formula {\sl for the same function $\ell$} that we
used for $\LL$.  This implies that $\LL$ and $\tilde\LL$ are identical functions. 

\section{The NLED/Particle-mechanics correspondence}\label{sec:SPD}

The starting point for section \ref{sec:SDH}  was the obvious fact that setting the magnetic field to zero 
in any Legendre pair of functions $\LL({\bf E},{\bf B})$ and $\HH({\bf D},{\bf B})$ yields functions $L(E)$ and $H(D)$, 
respectively, that are a Legendre pair {\sl when viewed as functions of ${\bf E}$ and ${\bf D}$, respectively}. We then showed that
this remains true if $L(E)$ and $H(D)$ are viewed as one-variable functions, and we explained how they are related
to the one-variable CH-functions $\ell$ and $\hh$ that determine  $\LL$ and $\HH$ for a  self-dual NLED. 

We now provide a different interpretation of the functions
$L(E)$ and $H(D)$, viewed as functions of ${\bf E}$ and ${\bf D}$, respectively. Rather than set to zero the magnetic field, as we did in section \ref{sec:SDH},  we replace the Euclidean 3-space with a flat 3-torus of 3-volume $v_3$,  and we truncate the Fourier expansion of fields 
on this 3-torus by setting all space derivatives of the 1-form potential to zero. 
We then have ${\bf B}= {\bf 0}$ but also ${\bf E} = -\dot{\bf A}$, where ${\bf A}(t)$ is the 3-vector potential, now a function only of the time coordinate $t$. The Lagrangian obtained by integrating over $T^3$ is therefore 
\be
\bb{L}(\boldsymbol{\nu})  = v_3 L(E) \, , \qquad \boldsymbol{\nu} := -\sqrt{v_3/m}\, {\bf E}\, , 
\ee
where $m$ is an arbitrary mass parameter needed to make $\nu$ dimensionless (for unit speed of light). 
The Hamiltonian (obtained by Legendre transform of $\bb{L}$) is 
\be
\bb{H}(\boldsymbol{\pi}) = v_3 H(D)\, , \qquad \boldsymbol{\pi} = - \sqrt{mv_3}\, {\bf D}\, , 
\ee
where $\boldsymbol{\pi}$ is the Legendre dual of $\boldsymbol{\nu}$. We may interpret $\bb{L}$ and $\bb{H}$ 
as the Lagrangian and Hamiltonian  for a point particle with velocity $\boldsymbol{\nu}$ and 
momentum $\boldsymbol{\pi}$ in a locally-Euclidean 3-space. 

For Maxwell we have ($\nu= |\boldsymbol{\nu}|$ and $\pi= |\boldsymbol{\pi}|$)
\be
\bb{L} = \frac12 v_3 E^2 = \frac12 m \nu^2 \, , \qquad \bb{H}= \frac12 v_3 D^2 = \frac{\pi^2}{2m}\, , 
\ee
which are the Lagrangian and Hamiltonian for a non-relativistic particle of mass $m$. 

For the Born-Infeld theory we may (since $m$ was an arbitrary mass parameter) set 
 \be 
T= m/v_3  \, ,
\ee
in which case 
\be
\bb{L}  = -m\sqrt{1- \nu^2} \, , \qquad \bb{H} = \sqrt{m^2 + \pi^2} \, , 
\ee 
which are the Lagrangian and Hamiltonian for a relativistic particle of mass $m$. As a consistency check, we observe that 
\be\label{nupi}
\tau = \frac12 T\nu^2 \, , \qquad \sigma = \frac12 T(\pi/m)^2\, , 
\ee
and the relations of  \eqref{sigtau} then imply
\be
\pi = \frac{m\nu}{\sqrt{1-\nu^2}}
\ee
as expected. 

There is a string-theory interpretation of this correspondence between the Born-Infeld field theory and the massive relativistic particle, because (as mentioned in the Introduction) it is implied by the T-duality relation between the D3-brane and D0-brane of Type II superstring theory. However the correspondence obtained above is more general because it applies to any 
self-dual NLED. For ModMaxBorn, for example, we find that  
\be
\bb{L}  = -m\sqrt{1- e^\gamma \nu^2} \, , \qquad \bb{H} = \sqrt{m^2 + e^{-\gamma}\pi^2} \, , 
\ee 
which are again the Lagrangian and Hamiltonian for a relativistic particle but with a ``modified light-speed'' of $e^{-2\gamma}$, which is subluminal for $\gamma>0$ and superluminal for $\gamma>1$. 
For this ``relativistic'' particle mechanics model, considered in isolation, we could redefine the ``speed of light'' to be $e^{-\gamma/2}$. However, as derived above, the particle mechanics model describes the `corner' of full self-dual NLED model for which $|{\bf B}|=0$ and 
$-{\bf E}$ is a space-independent but time-dependent 3-vector that we interpret as a particle velocity vector, and the speed of light is what it is in the full theory, i.e. unity. From this perspective, we should expect that the particle velocity can be superluminal only for an acausal NLED, and our ModMaxBorn example confirms this.

\section{Summary and outlook}

In any generalisation of an established physical theory, such as Maxwell electrodynamics, the question arises of which
features should be preserved and which may be discarded. The principal feature preserved by the Plebanski class 
of nonlinear electrodynamics is the canonical structure, and hence the number of degrees of freedom 
per space point. This means that small amplitude waves still have two distinct  polarisations, but these waves 
will typically interact with each other. In addition, they need not travel at light-speed, which leads to the possibility
of superluminal propagation in some backgrounds. This was initially investigated by considering  shock waves in 
generic smooth electromagnetic backgrounds, but equivalent  results are found by considering 
plane-wave perturbations of a generic constant uniform background, which can be viewed as a homogeneous
optical medium. 

For weak-field backgrounds (typically defined in relation to a Born scale introduced by interactions) the 
absence of superluminal propagation can be ensured by  imposing simple convexity conditions on the Lagrangian density. 
However, generic theories satisfying these conditions will still allow superluminal propagation for some strong-field 
backgrounds. The systematic study of this possibility dates back to a 2016 paper by Schellstede et al. \cite{Schellstede:2016zue}, whose results 
we have confirmed, and explored in the context of models proposed for a variety of phenomenological reasons over the past few decades \cite{Russo:2024kto}. One lesson from this work is that the simplest way to find a causal model is to choose one that 
is self-dual because weak-field causality implies strong-field causality (given the existence of a weak-field limit)  \cite{Russo:2024llm}.

Thus, one major reason for the study of self-dual NLED theories is that it is easy to separate the causal from the acausal cases.
In fact, this becomes even easier once it is appreciated that the Lagrangian density $\LL$ of any self-dual NLED theories (with a weak-field limit) can be constructed from a corresponding one-variable function $\ell$. This function, defined on a half-line, provides the boundary condition needed  to integrate the PDE that $\LL$ must satisfy for any self-dual theory; we have called this the Courant-Hilbert (CH) construction since the PDE and its solution can be found in \cite{C&H}. The causality conditions then reduce to simple constraints on the first and second derivatives of the function $\ell$ \cite{Russo:2024llm}. 

The initial aim of this paper was to  extend these results to the Hamiltonian formulation. Self-duality in this context is trivially ensured by restricting the Hamiltonian density 
$\HH$ to depend on duality-invariant variables, but now Lorentz invariance requires $\HH$ to satisfy a PDE, which (in appropriate variables) is formally the same as the one that $\LL$ must satisfy to ensure self-duality. This means that there is a CH construction for $\HH$ in terms of some other one-variable function $\hh$, also defined on a half-line, and $\{\LL,\HH\}$ is a Legendre-dual pair for any causal self-dual NLED.  We have shown that the one-variable functions $\{L,H\}$ defined by $L(\sqrt{2\tau}) = \ell(\tau)$ and $H(\sqrt{2\sigma})= \hh(\sigma)$ are also a Legendre-dual pair. This defines a 
correspondence between any causal self-dual NLED and a particle-mechanics model, with Born-Infeld corresponding to the massive relativistic point particle. 

The results just summarized also simplify the construction of $\HH$ from $\LL$, and {\it vice versa}, 
by reducing this problem to the Legendre transform of a one-variable function. However, a much greater
simplification is possible by taking advantage of a `dual' CH construction of $\LL$ from $\hh$. The fact that 
both $\LL$ and $\HH$ can be constructed from $\hh$ implies a very simple relation between them.
A procedure for finding $\HH$ given $\LL$, for example, is given in the one-line boxed equation 
\eqref{magic} of the Introduction. This is one of our main results, derived from an unexpected 
`duality'. 

Since $\LL$ and $\HH$ are so simply related, it is natural to suppose that the CH-functions $\ell$ and $\hh$ must also be related in a way that is simpler than via Legendre transform of the associated one-variable functions $(L,H)$. This is indeed true for some ``simple'' cases, such as Born-Infeld, but the general case requires consideration of what we have called ``$\Phi$-parity''. The variables $(U,V)$ are linear combinations of variables $(S,\Phi)$, with  $\Phi=\sqrt{S^2+P^2}$, and the $\Phi$-parity dual of $\LL(S,\Phi)$ is
$\hat\LL(S,\Phi) = \LL(S,-\Phi)$. The ``simple'' cases referred to above are those for which $\hat\LL=\LL$; i.e. the $\Phi$-parity invariant NLED theories. For these cases the CH functions $\{\ell,\hh\}$, both defined on a half-line, collectively define a single variable on a whole line; more precisely, they are related by the boxed equation \eqref{kinR} of section \ref{sec:ND}. 

Generically, $\hat\LL\ne\LL$ and we have a $\Phi$-parity pair of NLED theories with CH-functions $(\ell,\hh)$ and $(\hat\ell,\hat\hh)$, which are related in a similar way to \eqref{kinR}, but with a $\Phi$-parity twist; more precisely, the relations are those of the boxed equations \eqref{box1} and \eqref{box2} of section \ref{sec:ND}. In other words, the CH-function $\ell$ ($\hh$) of one self-dual NLED theory is simply related to the CH-function $\hat\hh$ ($\hat\ell$) of its $\Phi$-parity dual, and this reduces to the simple relation of 
\eqref{kinR} for the $\Phi$-parity invariant cases. 

A major theme of this paper has been that many interesting features of self-dual NLED theories are a corollary of simple features of their associated CH functions 
$\{\ell, \hh\}$. Examples are causality and conformal invariance, and the simple relation between Lagrangian and Hamiltonian densities summarised in the boxed equation \eqref{magic} of the Introduction. 

We have also shown how an alternative CH construction, again described by Courant and Hilbert, introduces a new
CH function, that in ``simple''  cases is (minus) the Legendre transform of the CH $\ell$-function. More generally, it is the Legendre transform of the $\ell$-function of a ``$\Phi$-parity'' dual theory. The ``simple'' cases are therefore those that are ``$\Phi$-parity'' invariant, and the simplest example is 
Born-Infeld. We have thus uncovered a new special 
property of Born-Infeld that may be of relevance in 
its applications, e.g. in string theory. 

Various other aspects of generic self-dual NLED theories deserve further investigation. It appears that only Born-Infeld is compatible with maximal (Minkowski spacetime)
supersymmetry, but the constraints of non-maximal supersymmetry are usually weaker. We have omitted coupling to electric and magnetic charges; their inclusion is of obvious interest, and we expect the results of \cite{Lechner:2022qhb}, for example, to be relevant to this. We also expect the CH construction of self-dual NLED theories to be useful in the exploration of NLED theories coupled to gravity. For example, the spacetime metric describing the analog of the Reissner-Nordstrom black hole might be expected to be invariant under an electromagnetic duality rotation of its parameters. We certainly 
expect causality to be a significant issue, and our previous result that the stress-energy tensors of causal NLED theories obey the same energy conditions as the Maxwell stress-energy tensor \cite{Russo:2024xnh} is 
an indication that results for Maxwell-Einstein will
generalize simply to self-dual NLED theories. 

To conclude, we remark that our Hamiltonian results can be applied directly to chiral 2-form electrodynamics in 6D Minkowski spacetime, for reasons spelled out in detail in \cite{Bandos:2020hgy}; essentially, one only has to re-interpret the variables $(u,v)$. As the 4D NLED theory is then a dimensional reduction from 6D we expect that
the 4D NLED causality conditions on $\hh$ will still apply, and may still be sufficient as well as necessary conditions for causality in 6D. This may be relevant to the recently investigated $T\bar T$-flows of 6D chiral 
2-form theories \cite{Ferko:2024zth}. 

\section*{Acknowledgements}
PKT has been partially supported by STFC consolidated grant ST/T000694/1. JGR acknowledges financial support from  grants 2021-SGR-249 (Generalitat de Catalunya) and a MINECO
grant PID2019-105614GB-C21.


\providecommand{\href}[2]{#2}\begingroup\raggedright\endgroup


\end{document}